\documentclass[preprint]{aastex631}

\shorttitle{Properties of star formation of the LMC as probed by YSOs}
\shortauthors{Kokusho et al.}
\graphicspath{{./}{figures/}}
\begin{document}

\title{Properties of star formation of the Large Magellanic Cloud as probed by young stellar objects}

\author{Takuma Kokusho}
\affiliation{Graduate School of Science, Nagoya University, Chikusa-ku, Nagoya, 464-8602, Japan}

\author{Hiroki Torii}
\affiliation{Graduate School of Science, Nagoya University, Chikusa-ku, Nagoya, 464-8602, Japan}

\author{Hidehiro Kaneda}
\affiliation{Graduate School of Science, Nagoya University, Chikusa-ku, Nagoya, 464-8602, Japan}

\author{Yasuo Fukui}
\affiliation{Graduate School of Science, Nagoya University, Chikusa-ku, Nagoya, 464-8602, Japan}

\author{Kengo Tachihara}
\affiliation{Graduate School of Science, Nagoya University, Chikusa-ku, Nagoya, 464-8602, Japan}



\begin{abstract} % 250 word limit

We perform a systematic study of evolutionary stages and stellar masses of young stellar objects (YSOs) in the Large Magellanic Cloud (LMC) to investigate properties of star formation of the galaxy. There are $4825$ sources in our YSO sample, which are constructed by combining the previous studies identifying YSOs in the LMC. Spectral energy distributions of the YSOs from optical to infrared wavelengths were fitted with a model consisting of stellar, polycyclic aromatic hydrocarbon and dust emissions. We utilize the stellar-to-dust luminosity ratios thus derived to study the evolutionary stages of the sources; younger YSOs are expected to show lower stellar-to-dust luminosity ratios. We find that most of the YSOs are associated with the interstellar gas across the galaxy, which are younger with more gases, suggesting that more recent star formation is associated with larger amounts of the interstellar medium (ISM). N157 shows a hint of higher stellar-to-dust luminosity ratios between active star-forming regions in the LMC, suggesting that recent star formation in N157 is possibly in later evolutionary stages. We also find that the stellar mass function tends to be bottom-heavy in supergiant shells (SGSs), indicating that gas compression by SGSs may be ineffective in compressing the ISM enough to trigger massive star formation. There is no significant difference in the stellar mass function between YSOs likely associated with the interface between colliding SGSs and those with a single SGS, suggesting that gas compression by collisions between SGSs may also be ineffective for massive star formation.

\end{abstract}



% Introduction
%%%%%%%%%%%%%%%%%%%%%%%%%%%%%%%%%%%%%%%%%%%%%%%%%%%%%%%%%%%%%%%%%%%%%%%%%
\section{Introduction} \label{sec:intro}

The Large Magellanic Cloud (LMC) is one of the nearest galaxies, located at a distance of ${\sim}50$~kpc \citep[e.g.][]{pie19}. Together with its proximity, a face-on view of the LMC \citep[$35^{\circ}$;][]{van01} enables us to study star formation across the galaxy in detail. In particular, the LMC is suitable for a study of star formation during the peak of cosmic star formation at the redshift of $z=1$--$2$ \citep{mad14}, as the metallicity of the LMC is similar to those of galaxies in this era \citep[$0.3$--$0.5Z_{\sun}$;][]{wes97}. Indeed the LMC hosts many active star-forming regions, such as N44, N157 (30 Doradus) and N159 \citep[e.g.][]{che09, che10,cro10}. The LMC has been known to harbor large shell structures bright in neutral and ionized gas emissions \citep{dav76, mea80, kim99}. These structures are called supershells, which are thought to have been formed by stellar feedback including radiation pressure and supernovae from multiple OB associations. Gas compression by the shocks plays an important role in the formation of molecular gases and stars. For instance, \citet{daw13} suggest that gas compression via interaction between the interstellar medium (ISM) and supershells promote molecular gas formation across the LMC disk. \citet{boo09} examined the spatial distributions of young stellar objects (YSOs) and supershells, to find that recent star formation preferentially occurs at peripheries of some supershells. On the other hand, it is suggested that the tidal interaction between the LMC and the Small Magellanic Cloud (SMC) induced the gas flow, triggering the massive star formation in the \ion{H}{1} ridge region encompassing the active star-forming regions N157 and N159 \citep{fuk17, fur19, mae21} and N44 \citep{tsu19, fur22}. As such, the LMC is likely to host star-forming regions possibly triggered by different mechanisms.

The nature of star formation in the LMC can be probed by YSOs which are protostars evolving into main-sequence stars, accompanied with infalling envelopes and accretion disks. Protostars heat their envelopes and disks consisting of gas and dust, producing an excess emission in the infrared (IR) wavelength over the protostar emission. Using the amplitudes of the IR excess, \citet{lad87} introduced a classification scheme for YSOs, namely Classes I, II and III, where the smaller number indicates younger objects. This classification accounts for the mid-IR spectral index, which becomes steeper for younger YSOs with larger dust envelopes and/or disks showing strong emission at longer wavelengths. Following the original classification by \citet{lad87}, \citet{and93} defined objects with almost no stellar emission as Class 0, which is likely to be in the earliest evolutionary phase of YSOs. The above evolutionary stages of YSOs can be used to study the evolution of star formation in galaxies; the presence of younger YSOs indicates that star formation more recently takes place there \citep[e.g.][]{boo09,kal18}.

The proximity and the nearly face-on view of the LMC mitigate the extinction due to the interstellar and intergalactic dust and the line-of-sight source confusion, respectively, facilitating a systematic study of YSOs across the galaxy. Various photometric surveys have been conducted for the LMC at multiwavelengths \citep[e.g.][]{hab99,kim99}. There are three large programs aiming to study YSOs in the LMC based on the IR survey of the galaxy; \citet{whi08} and \citet{gru09} independently established YSO catalogs of the LMC from the Surveying the Agents of a Galaxy's Evolution (SAGE) program with Spitzer \citep{mei06}. Although both studies utilized the same SAGE data, their classification schemes are somewhat different; \citet{whi08} adopted a strict YSO classification using the color-magnitude diagram based on theoretical YSO models, while \citet{gru09} adopted relatively relaxed selection rules in the color-magnitude diagram but carefully checked the morphology and the spectral energy distributions (SEDs) of their candidate sources. These two classifications accordingly have advantages and disadvantages. \citet{sea14} further identified YSOs in the LMC based on the far-IR photometry with Herschel to increase the number of Class 0/I objects in their catalog. Details of these YSO catalogs will be presented in Sect.~\ref{sec:catalog}.

In our study, we aim to investigate properties of star formation of the LMC based on the evolutionary stages and the spatial distributions of YSOs. We evaluate the evolutionary stages of YSOs from their stellar-to-dust luminosity ratios, which enable us to study the evolution of YSOs more quantitatively than the classical classification. Comparing the properties of the YSOs with tracers of the ISM in the LMC, we suggest star formation scenarios for the galaxy, which cover internal and external star formation triggers \citep[e.g.][]{boo09,fuk17}. Section~\ref{sec:data} describes our data set and classification scheme of the evolutionary stages of YSOs, the result of which is presented in Sect.~\ref{sec:res}. Our discussion on the properties of star formation of the LMC is presented in Sect.~\ref{sec:dis} and our conclusion in Sect.~\ref{sec:con}.

% Data
%%%%%%%%%%%%%%%%%%%%%%%%%%%%%%%%%%%%%%%%%%%%%%%%%%%%%%%%%%%%%%%%%%%%%%%%%
\section{Data and methods} \label{sec:data}

\subsection{YSO catalogs of the LMC} \label{sec:catalog}
The LMC was surveyed by Spitzer within the framework of the SAGE program \citep{mei06}. Using the SAGE data, \citet{whi08} and \citet{gru09} independently presented catalogs of YSOs in the LMC. \citet{whi08} used the SAGE point source catalog, in which source fluxes were measured with point-spread-function (PSF) fitting photometry. To select YSO candidates, they adopted color-magnitude criteria based on model SEDs of YSOs \citep{rob06}. They further examined the observed SEDs and Spitzer images of the YSO candidates to remove contamination by other sources such as background galaxies and produced the catalog consisting of $1197$ YSOs. On the other hand, \citet{gru09} performed aperture photometry to measure source fluxes and used simple color-magnitude criteria to extract YSO candidates. They also examined the observed SED and morphology of each source at multiwavelengths in detail to assess the nature of their YSO candidates. As a result, the YSO catalog containing $1385$ sources was established by \citet{gru09}, who reported that $379$ objects were matched within $1{\arcsec}$ with those in the catalog of \citet{whi08}. The color-magnitude criteria used in \citet{whi08} include fainter sources that are not covered by \citet{gru09}, while \citet{whi08} excluded sources with extended emission and those in crowded fields, which were covered by \citet{gru09}. As such, the two catalogs are complementary to each other, and thus we assume that the combination of the two catalogs offers us a more complete catalog of YSOs in the LMC. The differences in the properties of the YSOs between the two catalogs will be presented in Sect.~\ref{sec:res:luminosity}. Although there are some other studies identifying YSOs in the LMC \citep[e.g.][]{cau08, che09, che10, rom10, car12}, they focused on rather specific regions such as bright \ion{H}{2} regions, and methods to select YSOs are quite different from study to study. We thus combined the catalogs by \citet{whi08} and \citet{gru09} to cover the entire galaxy, using a matching distance of $1^{\arcsec}$, where the nearest sources are considered as matched objects when there are multiple sources within the matching distance. As a result, the combined catalog includes $2199$ YSOs.

Following Spitzer, Herschel observed the LMC in the far-IR wavelengths. Within the framework of the Herschel Inventory of the Agents of Galaxy Evolution (HERITAGE), \citet{mei13} established the far-IR point source catalog of the LMC. Based on this catalog, \citet{sea14} identified YSO candidates which were not detected significantly with Spitzer due to its lower sensitivity to the cold dust surrounding YSOs, i.e. Class 0/I. \citet{sea14} first excluded background galaxies by examining source morphologies and also excluded sources identified as non-YSOs in the previous studies. Then YSO candidates were selected by setting criteria such that sources are detected in more than two Herschel bands and in the Spitzer $24~{\mu}$m band, the latter being an indicator of the circumstellar dust of YSOs. As a result, $3518$ YSOs are included in the Herschel YSO catalog. We combined this with the above Spitzer catalog, using a matching distance of $5{\arcsec}$ which is usually used to match Herschel point sources with those from other point source catalogs \citep[e.g.][]{sea14}, to find that $892$ out of the $2199$ YSOs selected with Spitzer are successfully matched to those selected with Herschel. In total, there are $4825$ sources in the YSO sample used in our study.

We utilize dust-to-stellar luminosity ratios to evaluate the evolutionary stages of the sample YSOs (see Sect.~\ref{sec:sed}), and thus we need optical and/or near-IR photometry of the YSOs to estimate their stellar luminosities. The near-IR point source catalog of the LMC was produced by \citet{kat07} who measured the $J$, $H$ and $K$ band magnitudes of the LMC point sources with the InfraRed Survey Facility (IRSF). In the optical wavelength, \citet{zar04} established the point source catalog in the $U$, $B$, $V$ and $I$ bands from the Magellanic Clouds Photometric Survey (MCPS). We combined these two catalogs with our YSO sample, using matching distances of $1{\arcsec}$ and $5{\arcsec}$ for the YSOs detected by Spitzer and those detected only by Herschel, respectively.

\subsection{Photometry} \label{sec:phot}
Some of our sample YSOs lack photometry by Spitzer, although they are newly identified as YSOs with Herschel \citep{sea14}. Hence we performed photometry of such YSOs, using the Spitzer $3.6$, $4.5$, $5.6$, $8.0$ and $24~{\mu}$m band images published by the SAGE team \citep{mei06}. We adopted aperture photometry using the same aperture radius and sky annulus as those in \citet{gru09}. For the consistency check, we measured the flux densities of the YSOs which have the Spitzer measurement by \citet{whi08} and/or \citet{gru09} and confirm that our results are consistent with those of the two studies within ${\sim}20{\%}$, which is comparable to the discrepancy between the aperture and PSF-fitting photometries of the YSOs as reported by \citet{gru09}. We adopted the results of the PSF-fitting photometry of the Herschel data by \citet{mei13}, since the aperture photometry of the Herschel data can suffer severe contamination of background emission.

Some objects in our sample YSOs are not included in the near-IR point source catalog by \citet{kat07}, likely due to their detection threshold of signal-to-noise ratios ${\geq}~4$. We therefore performed aperture photometry for these sources, using the near-IR maps observed by IRSF\footnote{https://jvo.nao.ac.jp/portal/irsf.do}. Here the photometry radius of $3{\arcsec}$ and the sky annulus of $5{\arcsec}$--$7{\arcsec}$ were adopted. The former corresponds to two times the full-width at half maximum of the PSF of the IRSF instrument \citep{kat07}. We find that the near-IR flux densities thus derived are consistent within ${\sim}30{\%}$ with the original values in the catalog of \citet{kat07} who measured the flux densities with the PSF-fitting photometry.

\subsection{SED fits} \label{sec:sed}
The evolutionary stages of YSOs have been classified into Classes 0/I, II and III from earlier to later stages by their spectral indices in the mid-IR wavelength \citep{lad87}. \citet{rob06} introduced a different classification scheme based on their YSO model which uses radiation transfer codes and assumes disk geometries to infer the SEDs of YSOs at various evolutionary stages. They defined Stages I, II and III according to the envelope accretion rates and the stellar and disk masses in their model. In our study, we estimate the evolutionary stages of YSOs from the stellar-to-dust luminosity ratios, $L_\mathrm{star}/L_\mathrm{dust}$, derived from the SED fits. This procedure is similar to those by \citet{lad87} who assumes more dust emission and less stellar emission in younger YSOs. Our classification has an advantage over the conventional ones in that the evolutionary stages of YSOs are derived in a quantitative manner, which allows us to assess the nature of star formation of the LMC in more detail.

As described above, \citet{rob06} presented a detailed SED model of YSOs, which was later improved by \citet{rob17} with inclusion of cold dust emission in the far-IR wavelength and so on. Yet some limitations remain in the model; for instance, the model does not account for the emission from polycyclic aromatic hydrocarbons (PAHs) which are clearly detected in some YSOs \citep[e.g.][]{sea09,car12}. Hence we used a simple model in our SED fits; the stellar component was described with photospheric models by \citet{kur93}, where the metallicity of the LMC ([Fe/H]${\sim}-0.3$; \citealt{wes97}) was assumed. Stellar temperatures were allowed to vary between $3,500$ and $35,000$~K, while the surface gravity was fixed at ${\rm log}~g=4.0$ since this parameter has an insignificant effect on the model spectrum. To account for the foreground extinction, we applied the extinction curve of the LMC estimated by \citet{pei92} to the stellar component. The dust component was described by two- and three-temperature modified blackbody models for the YSOs not detected and detected by Herschel, respectively, where the emissivity power-law index was assumed to be $1.5$ \citep{oli19}. Relative strengths of the dust components are allowed to vary in our SED fits. When the fit with this model was not accepted with a $95{\%}$ confidence level, the PAH emission, as described by the \citet{dra07} model with a size distribution and an ionized fraction typical of the Galactic ISM, was added to the model. Even when the fit was accepted with the model without the PAH component, we added the PAH component to the model if the fit was improved significantly according to an F-test with a confidence level of $95{\%}$.

In fitting the SEDs of the sample YSOs, we applied a systematic error to each flux density. Based on the consistency check of our photometry (see Sect.~\ref{sec:phot}), we added $30{\%}$ and $20{\%}$ errors to the optical and near-IR bands and the Spitzer bands, respectively. The flux densities of the Herschel bands were measured by \citet{mei13} who noted that their PSF-fitting photometry possibly missed spatially extended emission. As they reported that their measurements were consistent with those by aperture photometry within a factor of $1.5$ or less, we added conservative systematic errors of $50{\%}$ to the Herschel flux densities.

% Results
%%%%%%%%%%%%%%%%%%%%%%%%%%%%%%%%%%%%%%%%%%%%%%%%%%%%%%%%%%%%%%%%%%%%%%%%%
\section{Results} \label{sec:res}

\subsection{Stellar and dust luminosities} \label{sec:res:luminosity}
\begin{figure}[t]
\includegraphics[width=\textwidth]{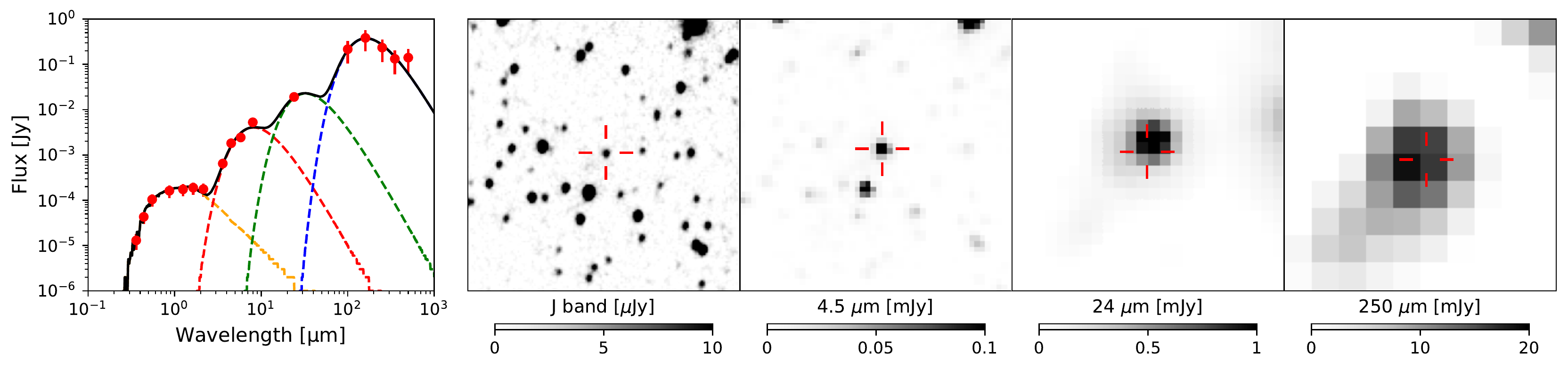}
\includegraphics[width=\textwidth]{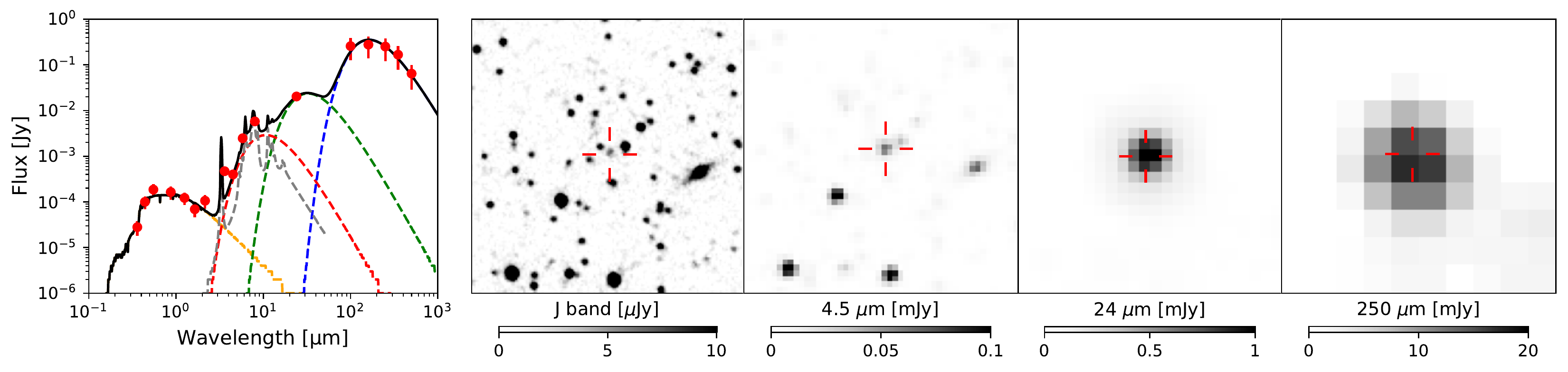}
\includegraphics[width=\textwidth]{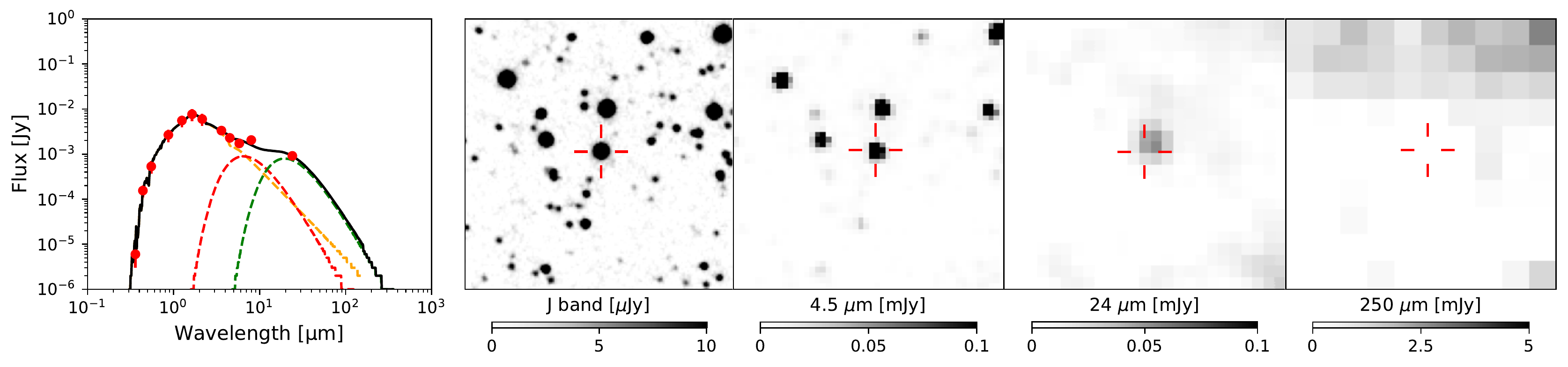}
\caption{Examples of the SEDs overlaid with the best-fit models and the multi-wavelength images in the $J$, $4.5~{\mu}$m, $24~{\mu}$m, and $250~{\mu}$m bands for well-fitted YSOs. In the left column, red points indicate observed data, while dotted orange, red, green, blue and gray lines indicate the best-fit stellar, hot dust, warm dust, cold dust and PAH components, respectively. Red crosses in the multi-wavelength images indicate the positions of the YSOs. \label{fig:sed}}
\end{figure}

\begin{figure}[t]
\centering
\includegraphics[width=0.6\textwidth]{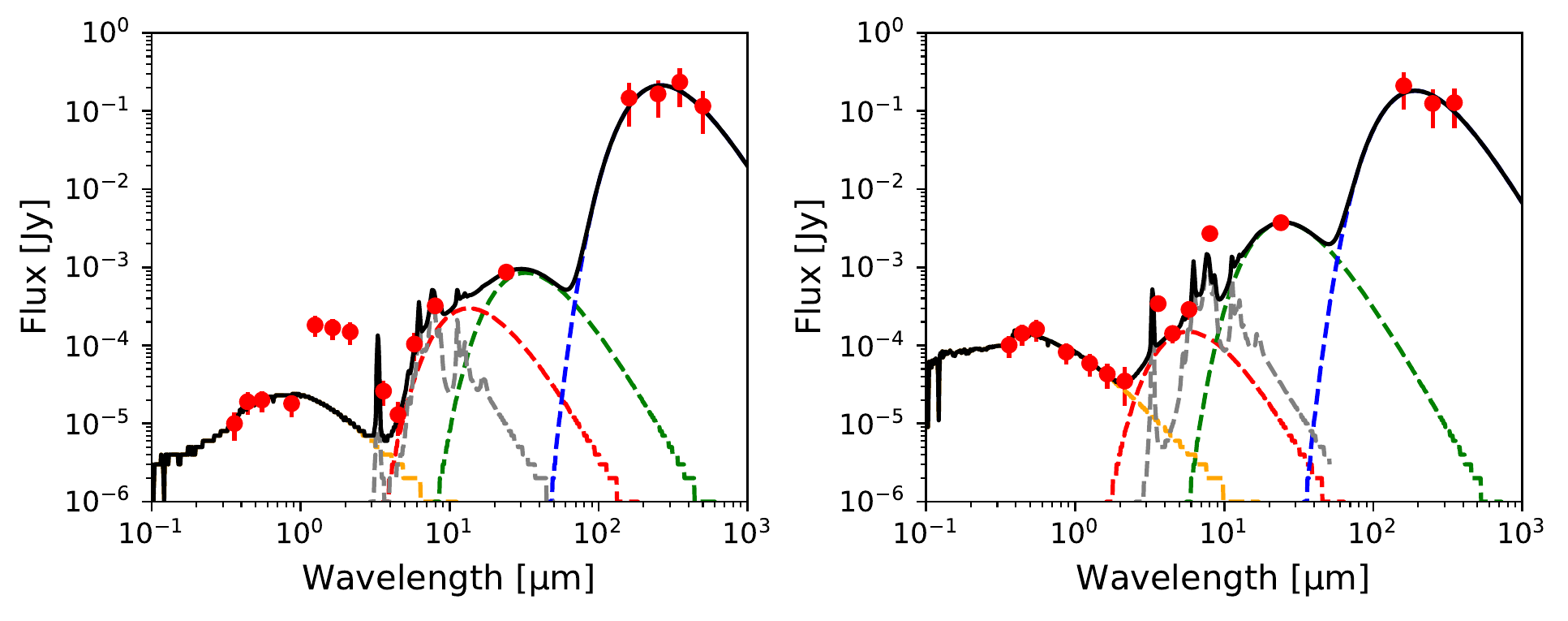}
\caption{Same as the left column of Fig.~\ref{fig:sed}, but for the SEDs not accepted in our study. \label{fig.sed_bad}}
\end{figure}

Figure~\ref{fig:sed} shows examples of the SED fits and the multi-wavelength images of our sample YSOs. We find that $4098$ out of the $4825$ sources are well-fitted YSOs,  i.e. their SED fits were accepted with a confidence level of $95{\%}$. Most of the rejected objects have no available MCPS and IRSF photometries or show inconsistent flux densities between the MCPS, IRSF and Spitzer bands, the latter indicating false matches between the point-source catalogs used in our study. In other cases, the model unsuccessfully reproduces the mid-IR emission for some objects, which is likely due to the presence of PAHs with size distributions and ionization degrees not typical of the Galactic ISM and/or strong silicate features. Figure~\ref{fig.sed_bad} shows examples of such SED fits not accepted in our study. We also find that $141$ well-fitted YSOs which have no optical counterparts show inconsistency between their model SEDs and the $V$ band limiting magnitude of the MCPS survey \citep{zar04}, indicating that $L_\mathrm{star}$ of these YSOs possibly has large uncertainties.

% SEDs

% bad SEDs

\begin{figure}[t]
\includegraphics[width=\textwidth]{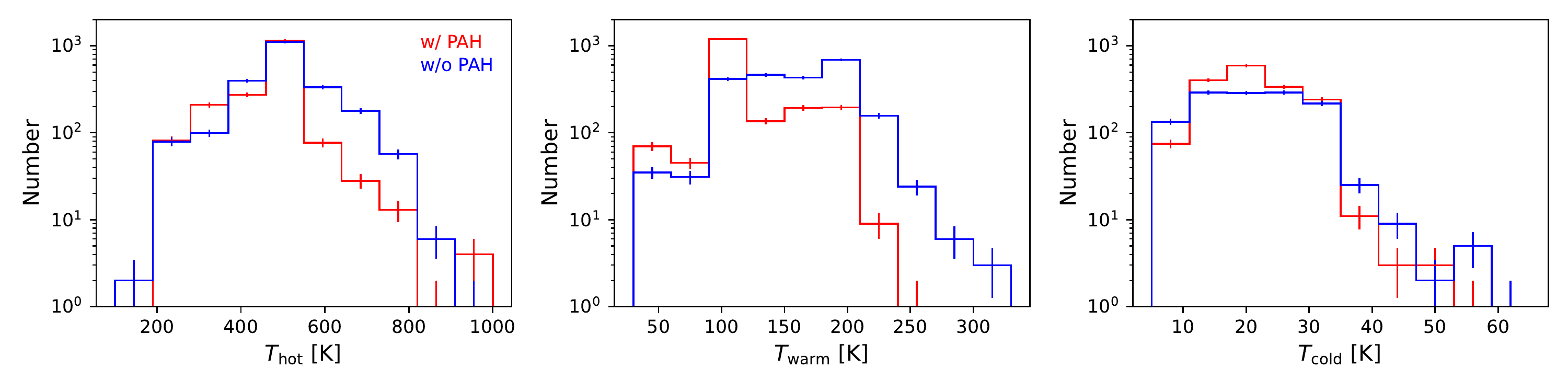}
\caption{Histograms of the dust temperatures for well-fitted YSOs. The dust temperatures are referred to as $T_\mathrm{hot}$, $T_\mathrm{warm}$ and $T_\mathrm{cold}$ from hotter to colder components. Red and blue histograms show the YSOs fitted with and without the PAH component, respectively. \label{fig:hist_T}}
\end{figure}

\begin{figure}[t]
\includegraphics[width=\textwidth]{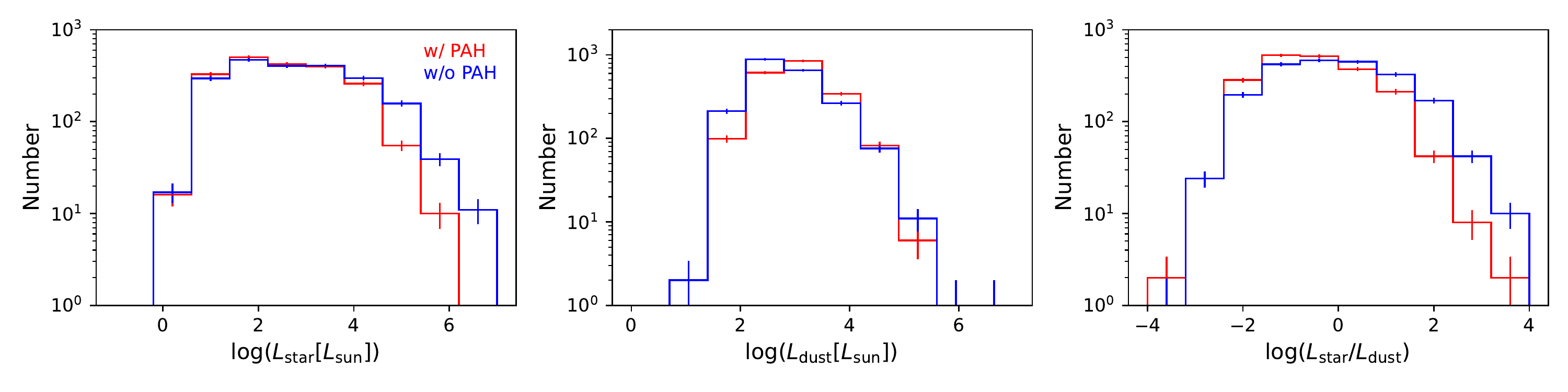}
\caption{Histograms of the stellar luminosities, dust luminosities and stellar-to-dust luminosity ratios for well-fitted YSOs. Red and blue histograms show the YSOs fitted with and without the PAH component, respectively. \label{fig:hist_L}}
\end{figure}

Figure~\ref{fig:hist_T} shows the dust temperatures obtained from the SED fits. Our sample YSOs have dust temperatures in a range between $10$ and $1,000$~K, which is consistent with the model prediction by \citet{dal98} who suggest that the dust temperature reaches ${\sim}1,000$~K in the inner disk, while dust cools down to ${\sim}10$~K in the outer disk. $L_\mathrm{star}$ and $L_\mathrm{dust}$ derived with the SED fits are shown in Fig.~\ref{fig:hist_L}, where $L_\mathrm{dust}$ is defined as the sum of the dust and PAH emissions. The left panel of the figure shows that most of the YSOs have $L_\mathrm{star}{\sim}10^1$--$10^6$~$L_\sun$. \citet{whi08} performed SED fits for $299$ YSOs, all of which are included in our sample, using the YSO model by \citet{rob06} to derive some parameters including $L_\mathrm{star}$, stellar masses and evolutionary stages. $287$ out of the $299$ YSOs correspond to well-fitted YSOs in our study. For these YSOs, we find that objects which have no optical counterparts show a random scatter of $0.4$~dex between $L_\mathrm{star}$ estimated by our study and \citet{whi08}, likely due to the different stellar models used in the two studies. On the other hand, YSOs which have optical counterparts show a larger random scatter of $1.1$~dex between the two samples, likely due to the fact that \citet{whi08} fitted the SEDs with only IR data. As we estimated $L_\mathrm{star}$ based on the optical and near-IR data for YSOs which have optical counterparts, our $L_\mathrm{star}$ is expected to be more reliable.

% histograms of temperature

% histograms of luminosity

In our sample YSOs, $1307$ objects are not detected by Herschel. We evaluated upper limits of luminosities of the cold dust component for these YSOs, using the median of the cold dust temperatures of the YSOs detected by Herschel of $21$~K and the $3{\sigma}$ upper limit of the $250$~$\mu$m band \citep{mei13}. We find that the inclusion of the upper limit of the cold dust luminosity increases $L_\mathrm{dust}$ by ${\sim}5\%$. Thus the non-detection in the far-IR bands is not likely to have significant effects on $L_\mathrm{dust}$. In a different vein, observed dust SEDs of YSOs can vary to some extent, depending on the inclination of the disks. To evaluate this effect on $L_\mathrm{dust}$, we used the model SEDs of \citet{rob06} who calculated SEDs at ten viewing angles for each model YSO. We integrated their dust SED over the wavelength range $10$--$1,000$~$\mu$m for each model YSO, to find that the dust flux thus derived changes by ${\sim}10{\%}$ depending on the viewing angle. Hence $L_\mathrm{dust}$ derived in our study potentially has such uncertainties.

% histograms of luminosity for W08 and GC09 YSOs
\begin{figure}[t]
\includegraphics[width=\textwidth]{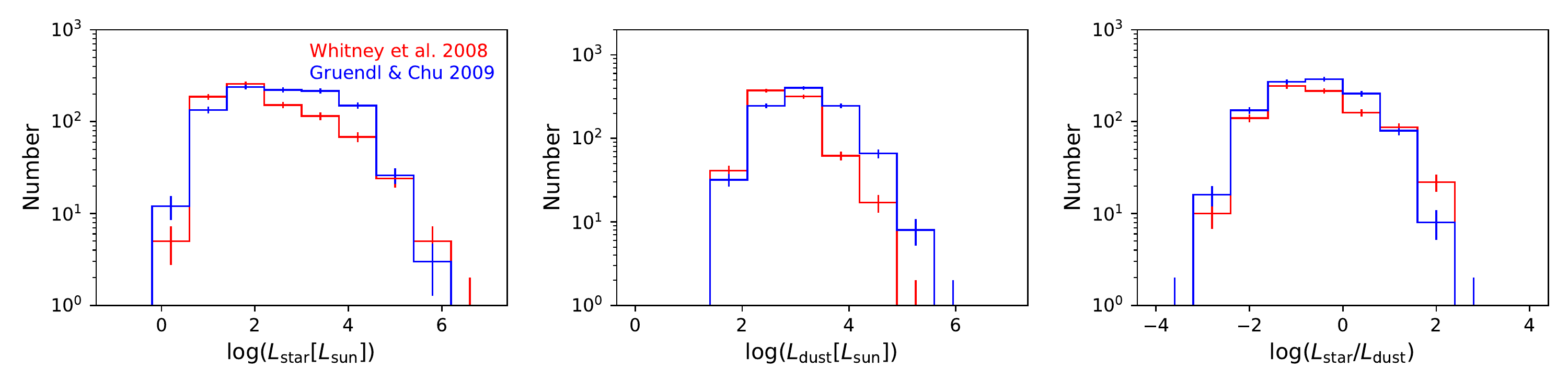}
\caption{Same as Fig.~\ref{fig:hist_L}, but for well-fitted YSOs selected either by \citet[red]{whi08} or by \citet[blue]{gru09}. \label{fig:hist_L_comp}}
\end{figure}

\begin{figure}[t]
\centering
\includegraphics[width=\textwidth]{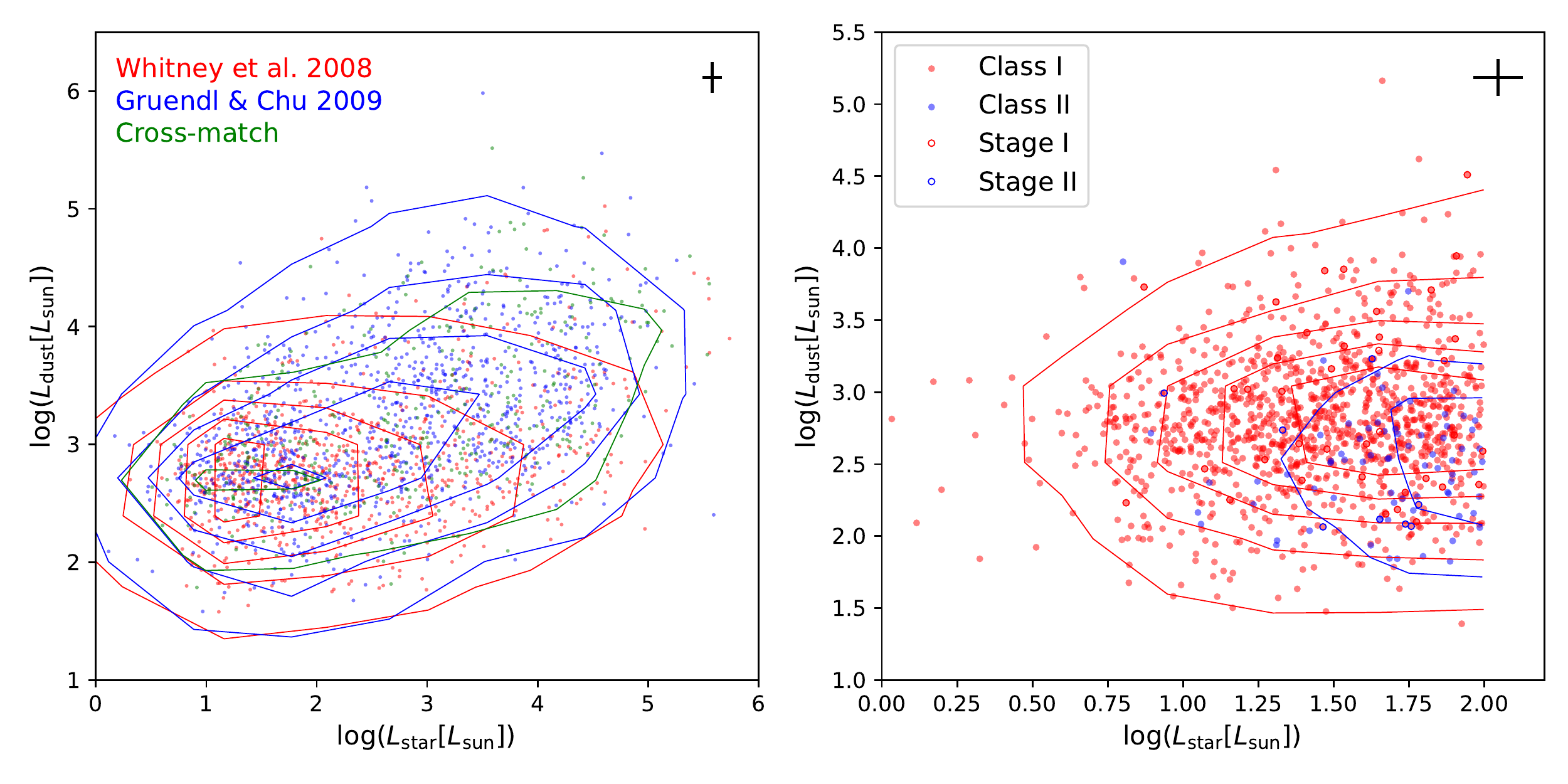}
\caption{Left: $L_\mathrm{dust}$ vs. $L_\mathrm{star}$ for well-fitted YSOs selected either by \citet[red]{whi08} or by \citet[blue]{gru09}. Cross-matched sources between the two catalogs are shown in green. Contours show densities of the data points. Typical errors along both axes are shown in the upper right. Right: same as the left panel, but for well-fitted YSOs with $L_\mathrm{star}$ less than $10^2~L_{\sun}$. Filled red and blue indicate YSOs classified as Classes I and II, respectively, while open red and blue indicate those classified as Stages I and II, respectively. Contours show densities of the data points for Classes I or II YSOs. \label{fig:plot_lumi}}
\end{figure}

We check the difference in $L_{\rm star}$ and $L_{\rm dust}$ of the YSOs between the catalogs of \citet{whi08} and \citet{gru09} in Fig.~\ref{fig:hist_L_comp} and the left panel of Fig.~\ref{fig:plot_lumi}. The figures show that YSOs selected only by \citet{gru09} tend to possess slightly higher $L_{\rm star}$ and $L_{\rm dust}$. This is likely due to the color-magnitude criteria adopted by \citet{whi08} which exclude brighter sources to suppress contamination of evolved stars. Thus we may miss massive YSOs showing higher $L_{\rm star}$ and $L_{\rm dust}$ in sparser fields where the source selection by \citet{whi08} is more accurate. However such massive objects tend to reside in crowded fields \citep[e.g.][]{gru09} where the source selection by \citet{gru09} is more accurate, and therefore the above difference in the source selection between the two catalogs is not likely to have significant effects on our study.

% Lstar vs. Ldust for W08 and GC09 YSOs and YSOs of different classes

To evaluate our classification of the evolutionary stages of YSOs, we derived the classification by \citet{lad87}, using the $K$ and $24~{\mu}$m band fluxes for well-fitted YSOs with $L_{\rm star}<10^2~L_{\sun}$, as the classification by \citet{lad87} is applicable mainly to low-mass objects. The right panel of Fig.~\ref{fig:plot_lumi} shows relations between $L_{\rm star}$ and $L_{\rm dust}$ for the low-$L_{\rm star}$ YSOs. The figure demonstrates that younger YSOs tend to show lower $L_{\rm star}$ and higher $L_{\rm dust}$. We find that the median values of log($L_\mathrm{star}/L_\mathrm{dust}$) are $-1.38$ and $-0.86$ for Classes I and II YSOs, respectively, showing that $L_\mathrm{star}/L_\mathrm{dust}$ is lower in younger YSOs as expected. In the same $L_{\rm star}$ range, the median values of  log($L_\mathrm{star}/L_\mathrm{dust}$) are $-1.42$ and $-0.53$ for Stages I and II YSOs classified by \citet{whi08}, respectively, again showing that $L_\mathrm{star}/L_\mathrm{dust}$ is lower in younger YSOs as expected. There is no Class III or Stage III YSOs in this $L_{\rm star}$ range as they are likely to have too faint disks to be included in the YSO catalogs used in our study. Overall, our classification of the evolutionary stages of YSOs is consistent with those in the literature.

$115$ well-fitted YSOs are detected in PAHs with the mid-IR spectroscopy \citep{jon17}. We find that $69$ out of the $115$ well-fitted YSOs call for the PAH component in our SED fits. Thus the detection rate of PAHs by our SED fits is $60{\%}$. Figure~\ref{fig:hist_L} shows that YSOs with non-detection of PAHs tend to have higher $L_\mathrm{star}$ and $L_\mathrm{star}/L_\mathrm{dust}$, indicating that PAHs may be destroyed and/or blown out from the central star due to the radiation field at later evolutionary stages. It is reported that low-mass YSOs tend to be lacking in PAHs with detection rates of only ${\sim}10{\%}$ \citep[e.g.][]{gee06}, which is significantly lower than the detection rate of PAHs for our sample YSOs. However our sample YSOs are biased to intermediate- to high-mass objects as described above, and therefore our results are likely to be applicable to only relatively massive YSOs. The key parameters obtained from our SED fits are summarized in Table~\ref{tab:sed}.

% parameters derived from SED fits
%\begin{longrotatetable}
\begin{deluxetable*}{ccccccccccc}
\tablecaption{Parameters for well-fitted YSOs \label{tab:sed}}
\tablewidth{0pt}
\tabletypesize{\scriptsize}
\tablehead{
\colhead{R.A. (J2000)} & \colhead{Decl. (J2000)} & \colhead{log($L_\mathrm{star}$)} & \colhead{log($L_\mathrm{dust}$)} & \colhead{log($L_\mathrm{hot}$)} & \colhead{log($L_\mathrm{warm}$)} & \colhead{log($L_\mathrm{cold}$)} & \colhead{log($L_\mathrm{PAH}$)} & \colhead{log($M_\mathrm{star}$)\tablenotemark{\tiny b}} & \colhead{$A_{V}$} & \colhead{Ref.\tablenotemark{\tiny c}} \\
\colhead{(deg)} & \colhead{(deg)} & \colhead{[$L_{\sun}$]} & \colhead{[$L_{\sun}$]} & \colhead{$[L_{\sun}$]} & \colhead{[$L_{\sun}$]} & \colhead{[$L_{\sun}$]} & \colhead{[$L_{\sun}$]} & \colhead{[$M_{\sun}$]} & \colhead{[mag]} & \colhead{}
}
\startdata
$71.82296$ & $-69.158400$ & $4.76{\pm}0.11$ & $4.21{\pm}0.05$ & $2.31{\pm}0.06$ & $4.08{\pm}0.06$ & $3.62{\pm}0.10$ & \nodata & $1.30{\pm}0.04$ & $6.8{\pm}0.6$ & 1, 2, 3 \\
$72.03664$ & $-68.705400$ & $2.27{\pm}0.10$ & $2.95{\pm}0.09$ & $2.09{\pm}0.05$ & $2.28{\pm}0.09$ & $2.76{\pm}0.13$ & \nodata & $0.58{\pm}0.02$ & $3.7{\pm}0.6$ & 1, 2, 3 \\
$72.15523$ & $-67.309700$ & $2.97{\pm}0.08$ & $3.10{\pm}0.15$ & $2.08{\pm}0.06$ & $2.77{\pm}0.08$ & $2.75{\pm}0.33$ & \nodata & $0.75{\pm}0.02$ & $18.6{\pm}0.0$ & 1, 2, 3 \\
$85.17992$ & $-70.186200$ & $3.14{\pm}0.21$ & $3.80{\pm}0.19$ & $1.96{\pm}0.17$ & $3.40{\pm}0.22$ & $3.53{\pm}0.30$ & $2.41{\pm}0.12$ & $0.79{\pm}0.05$ & $18.1{\pm}5.0$ & 1, 2, 3 \\ 
$85.18733$ & $-70.468580$ & $4.92{\pm}0.11$\tablenotemark{\tiny a} & $4.30{\pm}0.19$ & $3.03{\pm}0.05$ & $3.95{\pm}0.16$ & $4.00{\pm}0.36$ & \nodata & $1.36{\pm}0.04$ & $7.4{\pm}0.6$ & 1, 2, 3 \\
\enddata
\tablenotetext{a}{Model SED is not consistent with the $V$ band limiting magnitude \citep{zar04}.}
\tablenotetext{b}{$M_\mathrm{star}$ is estimated for YSOs which have $M_\mathrm{star}$ in the mass range $0.7$--$120$~$M_{\sun}$ (see text for details).}
\tablenotetext{c}{References identifying each YSO: (1) \citealt{whi08}; (2) \citealt{gru09}; (3) \citealt{sea14}.}
\tablecomments{Table~\ref{tab:sed} will be published in its entirety in the machine-readable format. A portion is shown here for guidance regarding its form and content.}
\end{deluxetable*}
%\end{longrotatetable}

\subsection{Evolutionary stages and spatial distributions of YSOs} \label{sec:res:map}
\begin{figure}[ht]
\includegraphics[width=\textwidth]{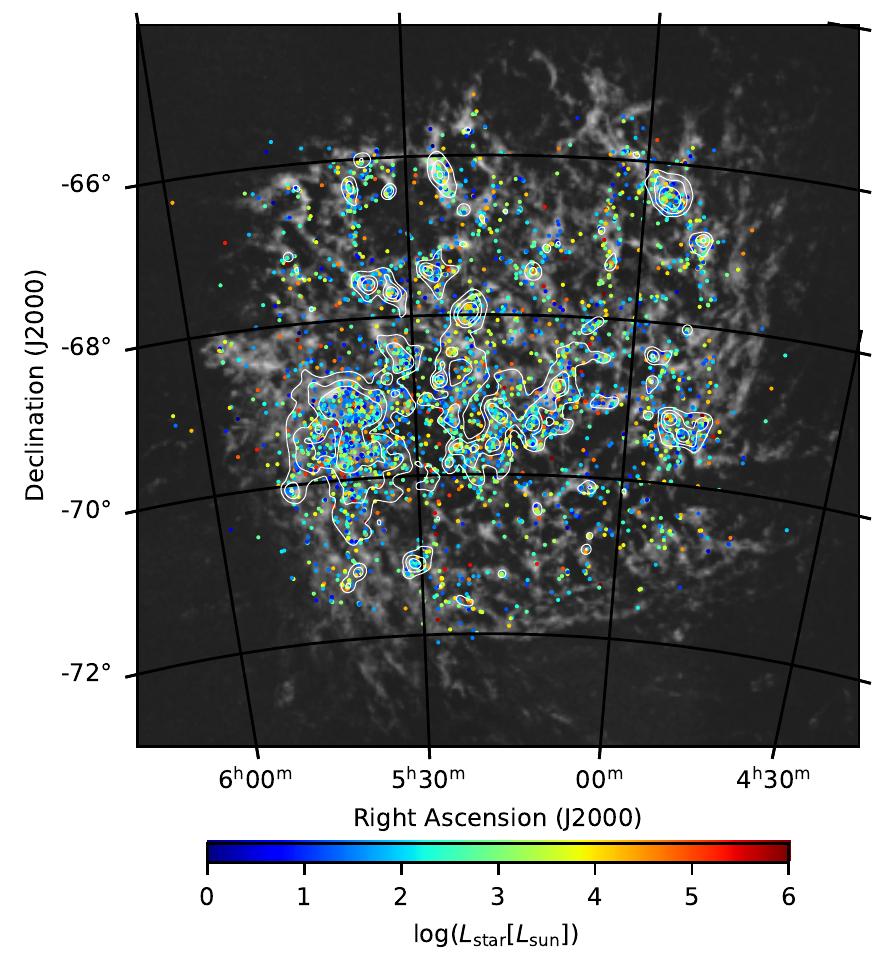}
\caption{Spatial distributions of well-fitted YSOs overlaid on the \ion{H}{1} peak brightness temperature map \citep{kim99}. The gray scale is in a range from $0$ to $100$~K in the brightness temperature. Data points are color-coded according to $L_\mathrm{star}$. White contours show the dust surface brightness map smoothed with a Gaussian kernel of $3{\arcmin}$ in sigma, which was calculated by using the results of \citet{gor14} (see text for details of the calculation). \label{fig:map_Ls}}
\end{figure}

\begin{figure}[ht]
\includegraphics[width=\textwidth]{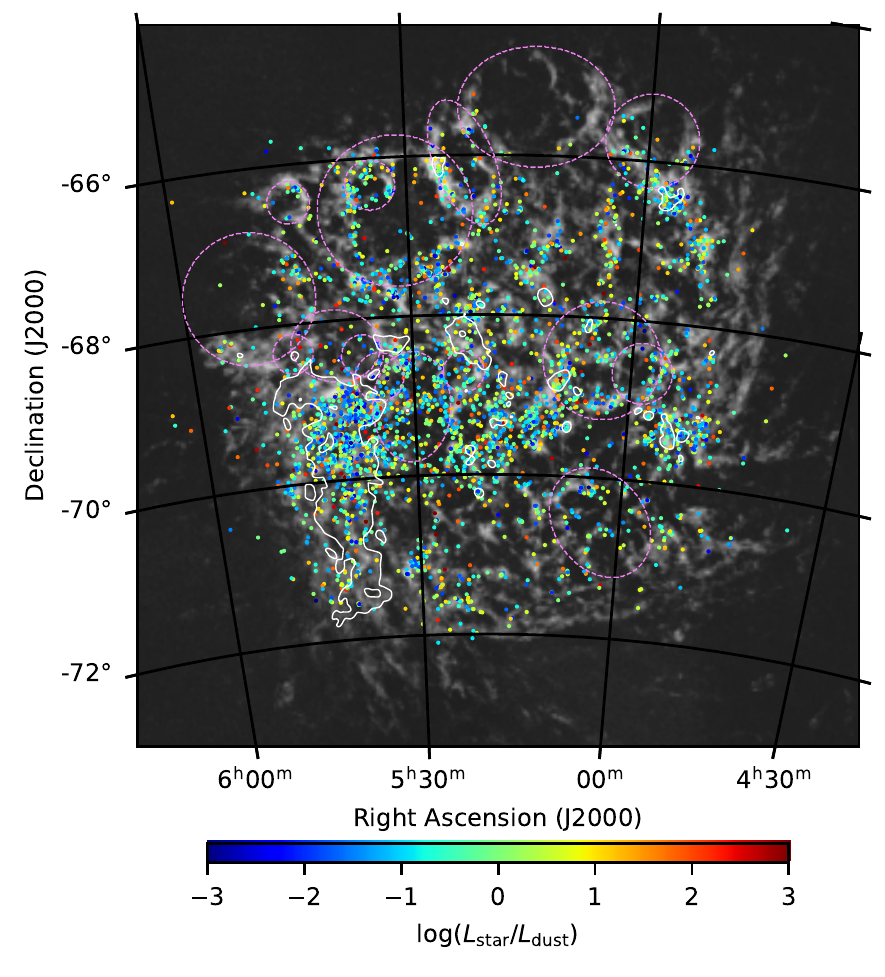}
\caption{Same as Fig.~\ref{fig:map_Ls}, but data points are color-coded according to $L_\mathrm{star}/L_\mathrm{dust}$. White contours and dotted violet lines show boundaries of the \ion{H}{1} ridge region and supergiant shells, respectively (see text for their definitions). \label{fig:map_Lsd}}
\end{figure}

\begin{figure}[ht]
\includegraphics[width=\textwidth]{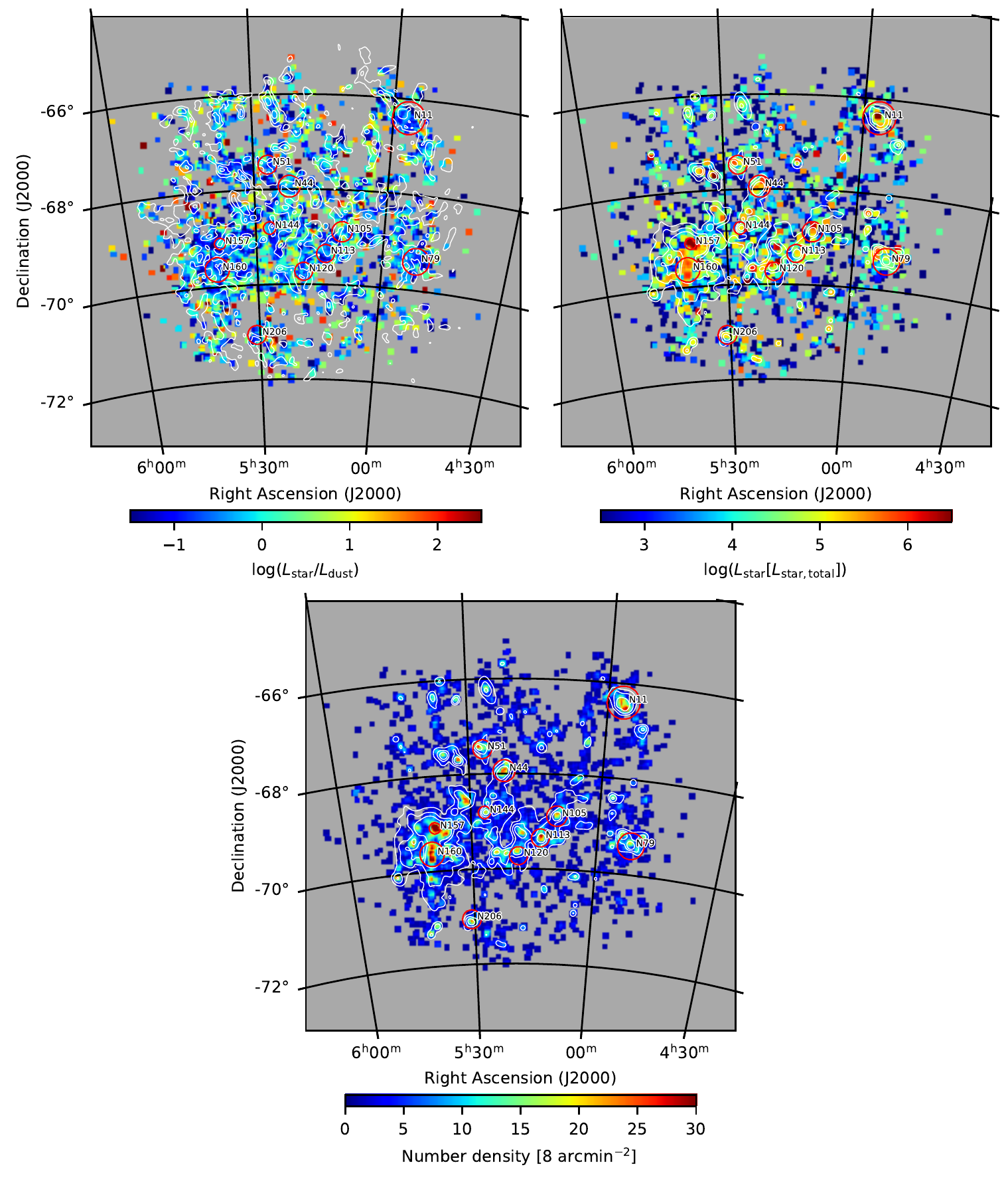}
\centering
\caption{The mean of $L_\mathrm{dust}/L_\mathrm{star}$ (top-left), the total $L_\mathrm{star}$ (top-right) and the number density (bottom) of well-fitted YSOs in every $8{\arcmin}{\times}8{\arcmin}$ region with a grid size of $40{\arcsec}$. Contours in the top-left panel show the \ion{H}{1} peak brightness temperature map \citep{kim99} smoothed with a Gaussian kernel of $2{\arcmin}$ in sigma. The contour levels are $30$, $50$ and $70$~K in the brightness temperature. Contours in the top-right and bottom panels are the same as those in Fig.~\ref{fig:map_Ls}. Red circles in all the panels indicate the locations of active star-forming regions. \label{fig:map_bin}}
\end{figure}

Figures~\ref{fig:map_Ls} and \ref{fig:map_Lsd} show the spatial distributions of the $4098$ well-fitted YSOs overlaid on the peak \ion{H}{1} brightness temperature map which traces dense \ion{H}{1} clouds along the line of sight \citep{kim99}. The data points in Figs.~\ref{fig:map_Ls} and \ref{fig:map_Lsd} are color-coded according to $L_\mathrm{star}$ and $L_\mathrm{star}/L_\mathrm{dust}$ derived from the SED fits, respectively. The figures show that the YSOs preferentially reside in the \ion{H}{1} gas as reported by previous studies \citep[e.g.][]{whi08,boo09}, while our study demonstrates this trend more clearly with much more YSOs across the entire galaxy. The top-left panel of Fig.~\ref{fig:map_bin} shows the mean of $L_\mathrm{dust}/L_\mathrm{star}$ of well-fitted YSOs in every $8{\arcmin}{\times}8{\arcmin}$ region. Figure~\ref{fig:map_Lsd} and the top-left panel of Fig.~\ref{fig:map_bin} show that YSOs with lower $L_\mathrm{star}/L_\mathrm{dust}$ tend to be associated with the \ion{H}{1} gas, while those with higher $L_\mathrm{star}/L_\mathrm{dust}$ are present outside the \ion{H}{1} gas, suggesting that recent star formation is associated with a large amount of the ISM. 

% Lstar map

% Lstar/Ldust map

% binned maps

\begin{figure}[t]
\centering
\includegraphics[width=0.6\textwidth]{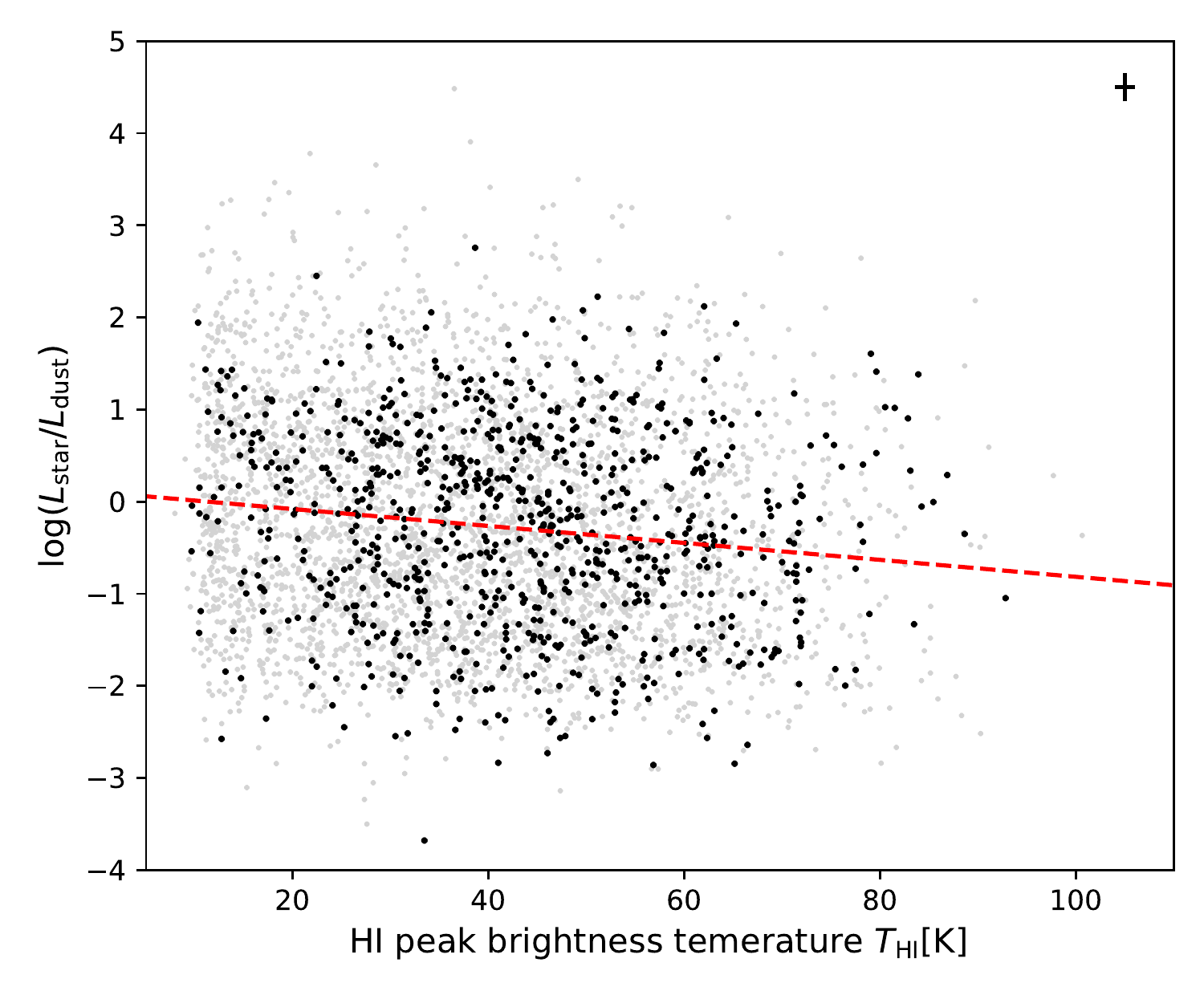}
\caption{$L_\mathrm{star}/L_\mathrm{dust}$ vs. the \ion{H}{1} peak temperature for well-fitted YSOs. Black and gray show Definite YSOs identified by \citet{gru09} and the other YSOs, respectively. Typical errors along both axes are shown in the upper right. The best linear fit to the data points is shown as dotted red lines, which is log~$L_\mathrm{star}/L_\mathrm{dust}=-0.01T_\mathrm{HI}+0.10$. \label{fig:plot_Lsd_HI}}
\end{figure}

Figure~\ref{fig:plot_Lsd_HI} shows relations between $L_\mathrm{star}/L_\mathrm{dust}$ and the peak \ion{H}{1} brightness temperature, where the linear correlation coefficient $R=-0.14$ and the probability of deriving the observed $R$ when the null hypothesis is true $p<0.01$. Hence the correlation is significant, and Fig.~\ref{fig:plot_Lsd_HI} demonstrates that recent star formation is indeed associated with a large amount of the ISM. Most likely contaminant sources in our YSO sample are evolved stars, the SEDs of which often mimic those of younger YSOs, i.e. objects with lower $L_\mathrm{star}/L_\mathrm{dust}$ \citep{jon17}. However evolved stars are usually present in isolated regions of low gas density \citep{jon15}, indicating that they are not likely to be the objects with lower $L_\mathrm{star}/L_\mathrm{dust}$ at higher \ion{H}{1} peak brightness temperature in Fig.~\ref{fig:plot_Lsd_HI}. Hence the anti-correlation in Fig.~\ref{fig:plot_Lsd_HI} is likely to be driven by YSOs. To suppress possible contamination of objects other than YSO in Fig.~\ref{fig:plot_Lsd_HI}, we adopted Definite YSOs identified by \citet{gru09}, a success rate of which is higher than $95{\%}$ according to the mid-IR spectroscopy \citep{sea09}. We find a significant correlation between $L_\mathrm{star}/L_\mathrm{dust}$ and the peak \ion{H}{1} brightness temperature with $R=-0.11$ and $p<0.01$ for Definite YSOs, suggesting that the correlation is indeed caused by YSOs. Definite YSOs also show a large scatter in $L_\mathrm{star}/L_\mathrm{dust}$ in Fig.~\ref{fig:plot_Lsd_HI}, suggesting that YSOs are likely to be in various evolutionary stages even in clouds of similar gas density.

% Lstar/Ldust vs HI

% Lstar/Ldust vs dust
\begin{figure}[t]
\centering
\includegraphics[width=0.6\textwidth]{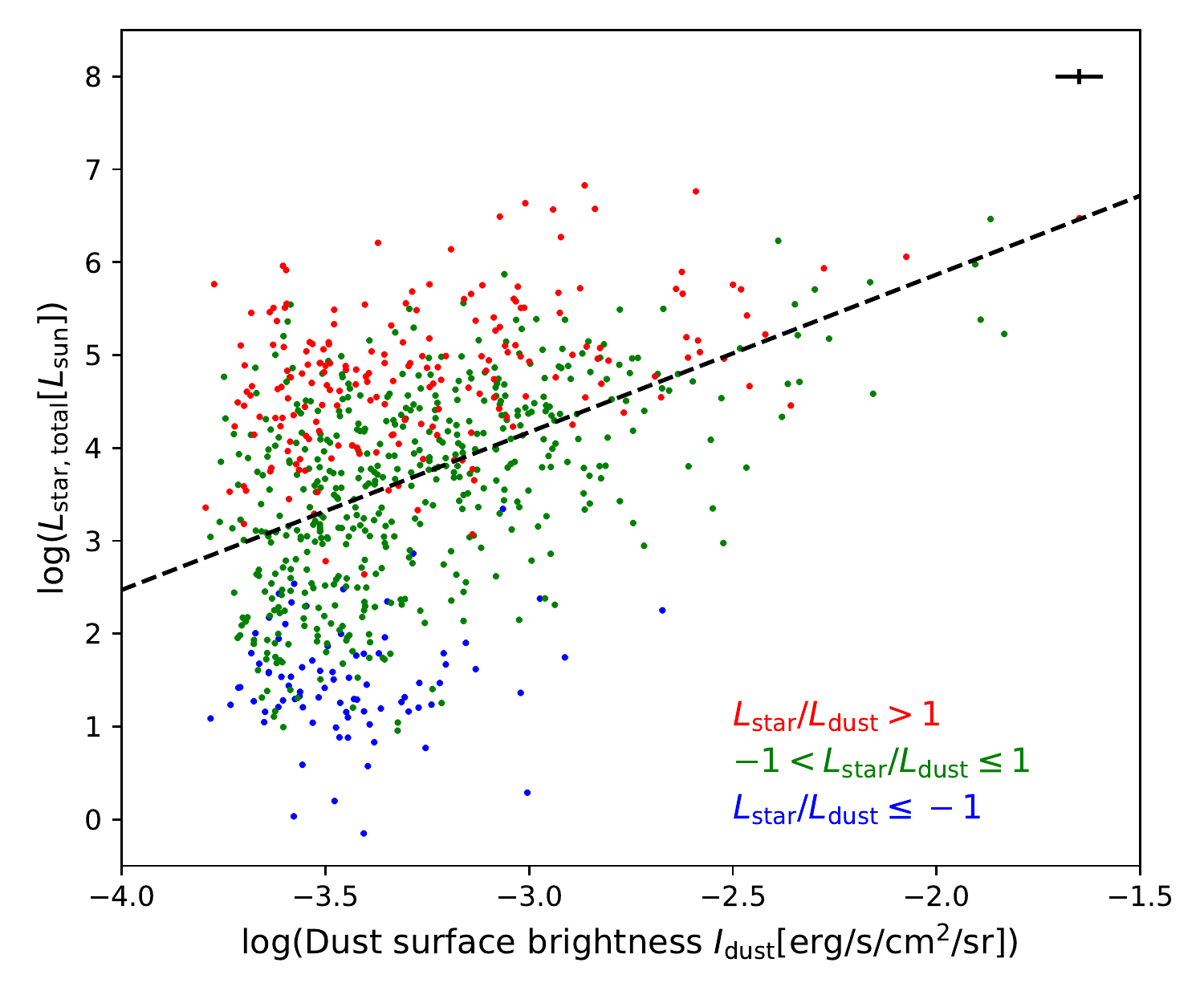}
\caption{Total $L_\mathrm{star}$ vs. the dust surface brightness for well-fitted YSOs. Data points are spatially sampled every $8{\arcmin}$. Red, green and blue show YSOs with $L_\mathrm{star}/L_\mathrm{dust}>1$, $-1<L_\mathrm{star}/L_\mathrm{dust}{\leq}1$ and $L_\mathrm{star}/L_\mathrm{dust}{\leq}-1$, respectively, where the mean of $L_\mathrm{star}/L_\mathrm{dust}$ in each $8{\arcmin}{\times}8{\arcmin}$ region is adopted. Typical errors along both axes are shown in the upper right. The best linear fit to the data points is shown as dotted black lines, which is log~$L_\mathrm{star, total}=1.70~$log~$I_\mathrm{dust}+9.26$. \label{fig:plot_Lsd_dust}}
\end{figure}

We also performed correlation analyses between $L_\mathrm{star}/L_\mathrm{dust}$ and the total gas column density, $N_\mathrm{H}$, and between $L_\mathrm{star}/L_\mathrm{dust}$ and the H$_2$ gas column density, $N_{\mathrm{H}_2}$, where $N_\mathrm{H}$ and $N_{\mathrm{H}_2}$ were estimated by \citet{tsu19} who considered both atomic and molecular hydrogen along each line of sight to derive $N_\mathrm{H}$. As a result, $R$ were estimated to be $-0.08$ and $-0.05$ with $p<0.01$ for $N_\mathrm{H}$ and $N_{\mathrm{H}_2}$, respectively, showing that recent star formation is indeed associated with a large amount of the ISM. Although these correlations are statistically significant, they are weaker than that estimated with the peak \ion{H}{1} brightness temperature. We therefore assume that the peak \ion{H}{1} brightness temperature is more suitable to trace the interstellar gas associated with YSOs than $N_\mathrm{H}$ and $N_{\mathrm{H}_2}$, since the latter can include significant amounts of background gases, which may obscure the correlation between $L_\mathrm{star}/L_\mathrm{dust}$ and the amount of the interstellar gas.

The top-right and bottom panels of Fig.~\ref{fig:map_bin} show the total $L_\mathrm{star}$ and the number density of well-fit YSOs, respectively. The former and the latter were calculated by summing up $L_\mathrm{star}$ and counting the number of YSOs in every $8{\arcmin}{\times}8{\arcmin}$ region, respectively. The figure reveals that active star-forming regions, such as N11, N113, N157 and N160 show higher total $L_\mathrm{star}$ and contain more YSOs, which is consistent with their higher star formation activities than other regions \citep[e.g.][]{cro10,car12}. Figure~\ref{fig:plot_Lsd_dust} shows relations between the total $L_\mathrm{star}$ and the dust surface brightness, where the latter was derived by using the results of \citet{gor14} who fitted far-IR SEDs with dust models across the LMC. We adopted their fits where they used a single temperature modified blackbody model with a broken power-law emissivity and integrated the best-fit model over the wavelength range $1$--$1,000$~$\mu$m to derive the dust surface brightness, which is an indicator of recent star formation activities \citep[e.g.][]{ken12}. The spatial scale of the dust map of \citet{gor14} is $56{\arcsec}$. We resampled their dust map to have the spatial scale of $8{\arcmin}$ for our analysis. $R$ was estimated to be $0.43$ with $p<0.01$ for the relation between the total $L_\mathrm{star}$ and the dust surface brightness. Hence the correlation is significant, indicating that many and/or massive YSOs are associated with active star-forming regions. Figure~\ref{fig:plot_Lsd_dust} also presents that older YSOs, i.e. those with higher $L_\mathrm{star}/L_\mathrm{dust}$, show higher total $L_\mathrm{star}$. This is likely due to the mass evolution of YSOs, which causes the scatter in the total $L_\mathrm{star}$ in Fig.~\ref{fig:plot_Lsd_dust}.

% Discussion
%%%%%%%%%%%%%%%%%%%%%%%%%%%%%%%%%%%%%%%%%%%%%%%%%%%%%%%%%%%%%%%%%%%%%%%%%
\section{Discussion} \label{sec:dis}

\subsection{Properties of overall star formation across the LMC} \label{sec:dis:sf}
In some star-forming regions in the LMC and our Galaxy, the ISM shows filamentary structures with widths down to $0.1$~pc \citep[e.g.][]{and17,fuk19,tok19}. This result is likely to support a picture for star formation suggested by numerical simulations which propose that multiple compression of the ISM by expanding supernova remnants and/or \ion{H}{2} regions is needed to form molecular cores and subsequently stars \citep{vai13,inu15,ino18}. Some mechanisms for gas compression in the LMC have been discussed in previous studies. For instance, \citet{kim99} identified a number of supershells in the LMC, which are formed by accumulation of shock waves of supernovae and expanding \ion{H}{2} regions, and the ISM is thought to be effectively compressed in the peripheries of supershells \citep[e.g.][]{daw13b}. Even smaller interstellar bubbles and supernovae can form new stars through accumulating the ISM, which results in multiple generation of star formation \citep[e.g.][]{che10}. In addition, gas compression induced by the tidal interaction between the LMC and the SMC is suggested for some star-forming regions in the LMC \citep{fuj90,fuk17,tsu20}. 

% parameters of SGSs
\begin{deluxetable*}{cccccccc}[t]
\tablecaption{Parameters for of the supergiant shells \label{tab:sgs}}
\tablewidth{0pt}
\tabletypesize{\small}
\tablehead{
\colhead{Name\tablenotemark{a}} & \colhead{R.A. (J2000)} & \colhead{Decl. (J2000)} & \colhead{Shell radius} & \colhead{Position angle} & \colhead{Number of YSOs\tablenotemark{b}} & \colhead{$R$\tablenotemark{c}} \\
\colhead{} & \colhead{} & \colhead{} & \colhead{(arcmin)} & \colhead{(deg)} & \colhead{} & \colhead{} & \colhead{}
}
\startdata
SGS2 & $04^\mathrm{h}58^\mathrm{m}30^\mathrm{s}$ & $-68{\degr}39{\arcmin}29{\arcsec}$ & $22.4{\times}22.4$ & $0$ & $54$ & $-0.14~(p=0.30)$ \\
SGS3 & $04^\mathrm{h}59^\mathrm{m}41^\mathrm{s}$ & $-65{\degr}44{\arcmin}43{\arcsec}$ & $35.0{\times}35.0$ & $0$ & $65$ & $0.06~(p=0.66)$ \\
SGS4 & $05^\mathrm{h}02^\mathrm{m}51^\mathrm{s}$ & $-70{\degr}33{\arcmin}15{\arcsec}$ & $34.0{\times}34.0$ & $36$ & $79$ & $0.00~(p=0.98)$ \\
SGS5 & $05^\mathrm{h}04^\mathrm{m}08^\mathrm{s}$ & $-68{\degr}31{\arcmin}55{\arcsec}$ & $43.9{\times}43.9$ & $0$ & $238$ & $0.02~(p=0.71)$ \\
SGS6 & $05^\mathrm{h}13^\mathrm{m}58^\mathrm{s}$ & $-65{\degr}23{\arcmin}27{\arcsec}$ & $45.0{\times}45.0$ & $90$ & $23$ & $-0.21~(p=0.35)$ \\
SGS7 & $05^\mathrm{h}22^\mathrm{m}42^\mathrm{s}$ & $-66{\degr}05{\arcmin}38{\arcsec}$ & $22.4{\times}22.4$ & $21$ & $91$ & $-0.05~(p=0.67)$ \\
SGS8 & $05^\mathrm{h}23^\mathrm{m}05^\mathrm{s}$ & $-68{\degr}42{\arcmin}49{\arcsec}$ & $15.0{\times}15.0$ & $0$ & $39$ & $-0.11~(p=0.52)$ \\
SGS11 & $05^\mathrm{h}31^\mathrm{m}33^\mathrm{s}$ & $-66{\degr}40{\arcmin}28{\arcsec}$ & $56.6{\times}56.6$ & $85$ & $307$ & $0.00~(p=1.00)$ \\
SGS12 & $05^\mathrm{h}30^\mathrm{m}26^\mathrm{s}$ & $-69{\degr}07{\arcmin}56{\arcsec}$ & $29.0{\times}29.0$ & $0$ & $307$ & $-0.12~(p=0.03)$ \\
SGS14 & $05^\mathrm{h}34^\mathrm{m}33^\mathrm{s}$ & $-66{\degr}20{\arcmin}15{\arcsec}$ & $18.5{\times}18.5$ & $0$ & $69$ & $0.02~(p=0.85)$ \\
SGS15 & $05^\mathrm{h}34^\mathrm{m}44^\mathrm{s}$ & $-68{\degr}44{\arcmin}36{\arcsec}$ & $17.9{\times}17.9$ & $0$ & $55$ & $-0.08~(p=0.57)$ \\
SGS16 & $05^\mathrm{h}36^\mathrm{m}54^\mathrm{s}$ & $-68{\degr}27{\arcmin}46{\arcsec}$ & $15.3{\times}15.3$ & $0$ & $24$ & $0.05~(p=0.81)$ \\
SGS17 & $05^\mathrm{h}40^\mathrm{m}26^\mathrm{s}$ & $-68{\degr}18{\arcmin}58{\arcsec}$ & $26.7{\times}26.7$ & $85$ & $46$ & $-0.11~(p=0.47)$ \\
SGS21 & $05^\mathrm{h}44^\mathrm{m}53^\mathrm{s}$ & $-66{\degr}28{\arcmin}29{\arcsec}$ & $15.6{\times}15.6$ & $0$ & $18$ & $0.12~(p=0.64)$ \\
SGS22 & $05^\mathrm{h}46^\mathrm{m}16^\mathrm{s}$ & $-68{\degr}22{\arcmin}02{\arcsec}$ & $14.7{\times}14.7$ & $0$ & $11$ & $-0.62~(p=0.04)$ \\
SGS23 & $05^\mathrm{h}51^\mathrm{m}16^\mathrm{s}$ & $-67{\degr}37{\arcmin}18{\arcsec}$ & $49.6{\times}49.6$ & $0$ & $73$ & $-0.05~(p=0.69)$ \\
\enddata
\tablenotetext{a}{Original name by \citet{kim99}.}
\tablenotetext{b}{The number of YSOs present inside the shell.}
\tablenotetext{c}{Correlation coefficients together with their $p$ calculated for the relations between $L_\mathrm{star}/L_\mathrm{dust}$ and the distances from the center for the YSOs present inside the shell.}
\end{deluxetable*}

The top-right and bottom panels of Fig.~\ref{fig:map_bin} show that many and/or massive YSOs tend to be present in active star-forming regions, such as N11, N113, N157 and N160. In particular, N157 shows the highest total $L_\mathrm{star}$ and number density of the YSOs, which is consistent with the fact that N157 is the most massive star-forming region in the LMC \citep[e.g.][]{cro10}. Star formation in N157 and N159 is thought to have been induced by the tidal interaction between the LMC and the SMC \citep{fuk17,fuk19,tok19}, while that in N11 may have been induced by supershells \citep{hat06}. N113 is associated with the stellar bar region where gas compression by stellar feedback is likely to take place \citep{oli06}.

% histograms of Lstar/Ldust
\begin{figure}[t]
\centering
\includegraphics[width=0.5\textwidth]{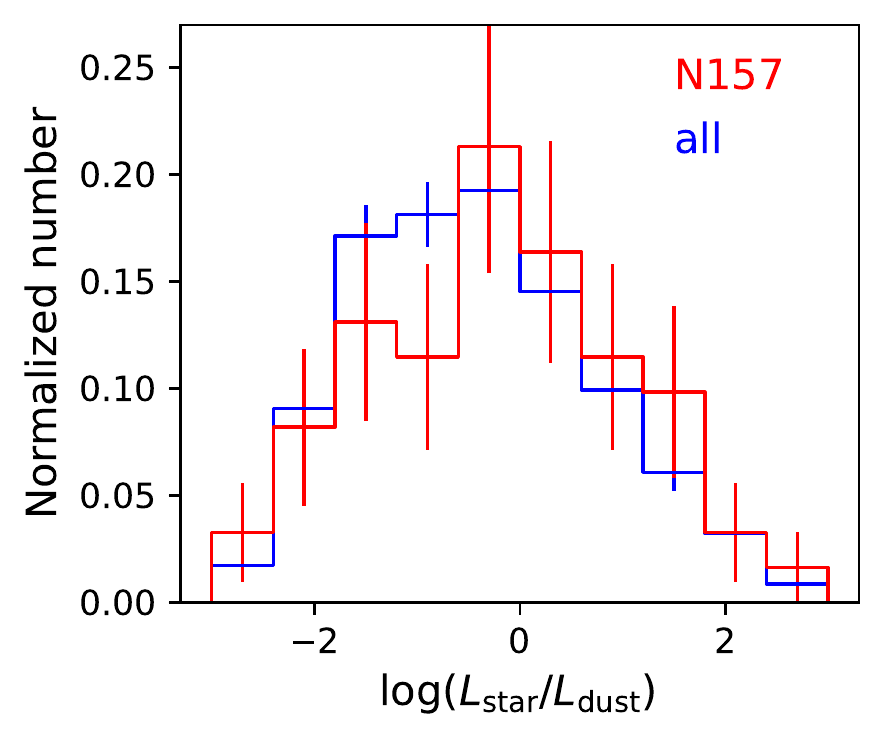}
\caption{Normalized histograms of $L_\mathrm{star}/L_\mathrm{dust}$ for well-fitted YSOs. Red and blue histograms show YSOs associated with N157 and all the star-forming regions defined in Fig.~\ref{fig:map_bin}, respectively. \label{fig:hist_Lsd}}
\end{figure}

We investigate the evolutionary stages of YSOs in the active star-forming regions shown in Fig.~\ref{fig:map_bin}, where the locations and the extents of the star-forming regions are defined by previous studies \citep{car12,lop14,och17}. We examined the distribution of $L_\mathrm{star}/L_\mathrm{dust}$ of YSOs in each star-forming region, to find that N157 tends to show $L_\mathrm{star}/L_\mathrm{dust}$ different from the average of all the star-forming regions as shown in Fig.~\ref{fig:hist_Lsd}; the difference in the distributions of $L_\mathrm{star}/L_\mathrm{dust}$ is significant according to a Kolmogorov-Smirnov (KS) test with the probabilities that the two samples are from the same population $p_\mathrm{KS}$ of $<0.01$. Figure~\ref{fig:hist_Lsd} represents that YSOs in N157 tend to show systematically higher $L_\mathrm{star}/L_\mathrm{dust}$, suggesting that recent star formation, i.e. mass evolution, of N157 may be in later evolutionary stages, which is consistent with the fact that N157 harbors massive clusters \citep[e.g.][]{cro10}. We find that five out of $64$ YSOs in N157 are located within \ion{H}{2} regions of O stars \citep{bon09}, assuming a typical radius of \ion{H}{2} regions by O stars of $1.5$~pc \citep{tie05}. This indicates that the feedback from massive stars may not significantly contribute to the recent star formation in N157. However we may miss fainter YSOs in such a crowded star-forming region due to the completeness limit of the catalog by \citet{whi08}, and the above scenario should be verified with a more complete YSO catalog.

%Other star-forming regions such as N79 are thought to be in earlier evolutionary stages and expected to evolve into massive clusters in the future \citep{och17}.

We also investigate the stellar mass functions of the YSOs separately for the supershell, \ion{H}{1} ridge and other field regions which are likely to have different mechanisms for gas compression. We here consider supergiant shells (SGSs) originally identified by \citet{kim99} and later examined in detail by \citet{boo08} who found that some SGSs in \citet{kim99} are false identifications. We excluded SGSs largely overlapped with the \ion{H}{1} ridge region (see below) from our analysis and enlarged the shell radius measured by \citet{kim99} with a factor of $1$--$2$ to cover the entire SGSs. The parameters of the SGSs are summarized in Table~\ref{tab:sgs} and their positions are shown in Fig.~\ref{fig:map_Lsd}. \citet{tsu20} analyzed the tidally-driven gas flow by decomposing the hydrogen gases of the LMC into three velocity components, to find that the intermediate velocity component in a range from $-30$ to $-10$~km~s$^{-1}$ shows bridge features indicative of dynamical interaction between the gas flow and the LMC disk. We defined the \ion{H}{1} ridge region as that with the gas column density of the intermediate velocity component higher than $1{\times}10^{21}$~cm$^{-2}$, which is shown in Fig.~\ref{fig:map_Lsd}. Figure~\ref{fig:map_Lsd} demonstrates that this definition reasonably traces star-forming regions where star formation induced by the gas flow is suggested \citep{fuk17,fur19}. We selected field regions from those which do not overlap with the SGS or \ion{H}{1} ridge regions thus defined.

\begin{figure}[t]
\centering
\includegraphics[width=\textwidth]{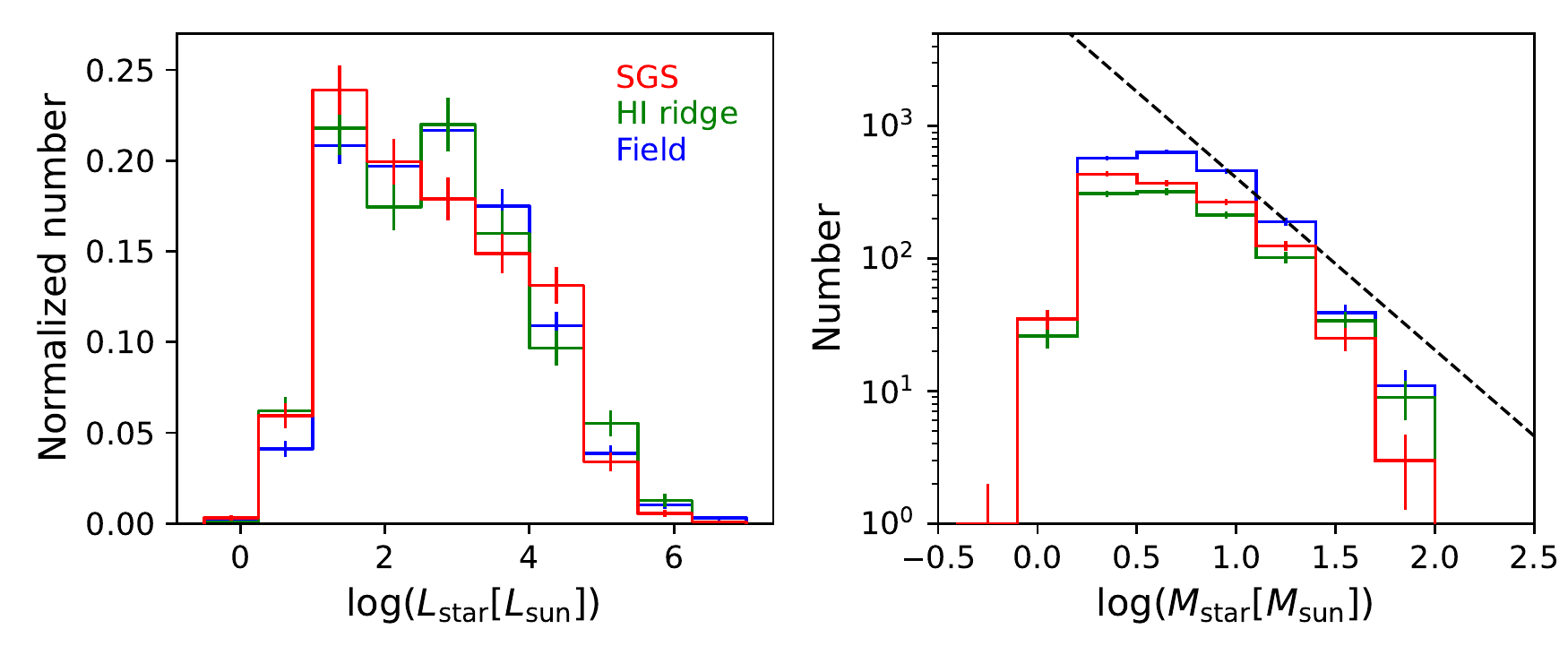}
\caption{Normalized histograms of the stellar luminosities (left) and histograms of the stellar masses (right) for well-fitted YSOs. Red, green and blue histograms show YSOs associated with the SGS, \ion{H}{1} ridge and field regions, respectively. Black dotted line in the right panel shows the Kroupa initial mass function \citep{kro01} \label{fig:hist_all}}
\end{figure}

The left panel of Fig.~\ref{fig:hist_all} shows the histograms of $L_\mathrm{star}$ for the three regions, demonstrating that YSOs in the SGS region are relatively biased toward low-$L_\mathrm{star}$ objects compared to the other regions. We estimated stellar masses, $M_\mathrm{star}$, of the YSOs using the pre-main sequence evolutionary tracks calculated by \citet{hae19}. We fitted the relations between $L_\mathrm{star}$ and $M_\mathrm{star}$ for their stellar models of ages between $10^5$ and $10^7$~yr and $M_\mathrm{star}$ between $0.7$ and $120~M_\sun$ with a cubic function. The fitted curve reasonably traces this relation, and $M_\mathrm{star}$ thus derived has errors up to a factor of $3$ due to a range of possible evolutionary tracks for each $L_\mathrm{star}$. $M_\mathrm{star}$ of each YSO is derived from their $L_\mathrm{star}$ as shown in the right panel of Fig.~\ref{fig:hist_all}. The YSO catalogs used in our study miss high- and low-mass YSOs, likely due to faster evolution of massive objects and detection limits, respectively \citep{whi08,car12}.

% histograms of Lstar & Mstar

In the right panel of Fig.~\ref{fig:hist_all}, the YSOs in the SGS region show a hint of a steeper slope toward the high-mass end than the other regions. Indeed a KS test shows that the distribution of $M_\mathrm{star}$ of the SGS region is significantly different from those of the \ion{H}{1} ridge and field regions with $p_\mathrm{KS}$ of $0.06$ and $<0.01$, respectively. \citet{ino13} suggest that collision between molecular clouds can trigger the formation of massive stars. \citet{ino18} and \citet{fuk21b} further claim that the gas column density higher than $1{\times}10^{22}$~cm$^{-2}$ is needed to form massive stars in the post-shock region. We speculate that it may be difficult for interaction between shocks and the ISM in the SGS region to produce such a high column density. Indeed the Galactic supershells are thought to be ineffective in triggering massive star formation \citep{daw11}. In addition, the shock of SGS whose velocity is up to $50$~km~s$^{-1}$ \citep{boo09} may destroy clouds in the shell regions to some extent, suppressing massive star formation there \citep{inu15,nto17}. On the other hand, collisions between clouds can effectively produce column densities high enough to form massive stars \citep{ino18}. This is applicable to the \ion{H}{1} ridge and field regions; the LMC clouds are likely to interact with the tidally-driven gas flow in the \ion{H}{1} ridge region \citep{fuk17,tsu20}, while clouds may interact with each other in the field region as in our Galaxy \citep[e.g.][]{fuk21a,eno21}.

The right panel of Fig.~\ref{fig:hist_all} also shows that the distribution of $M_\mathrm{star}$ in the \ion{H}{1} ridge region tends to be flatter toward the high-mass end, possibly indicating that collisions between clouds induced by the tidally-driven gas flow may be effective for massive star formation \citep{fuk17,ino18, tsu20}. More detailed analysis on massive star formation by using the spatial distribution and $M_\mathrm{star}$ of O stars in the LMC will be reported in a future paper.

\subsection{Properties of local star formation in the SGS region} \label{sec:dis:ss}
SGSs are thought to have swept up the ISM by their expanding shell to trigger star formation \citep[e.g.][]{daw13,fuj14}. Although recent star formation is assumed to be associated with some SGSs in the LMC \citep{boo09}, previous studies mainly rely on the positional coincidence of young massive stars and the peripheries of SGSs to conclude that SGSs induce star formation. As discussed earlier, other mechanisms can enhance star formation around SGSs by chance. We therefore assess triggered star formation in SGSs, using the evolutionary stages of our sample YSOs.

\begin{figure}[t]
\includegraphics[width=\textwidth]{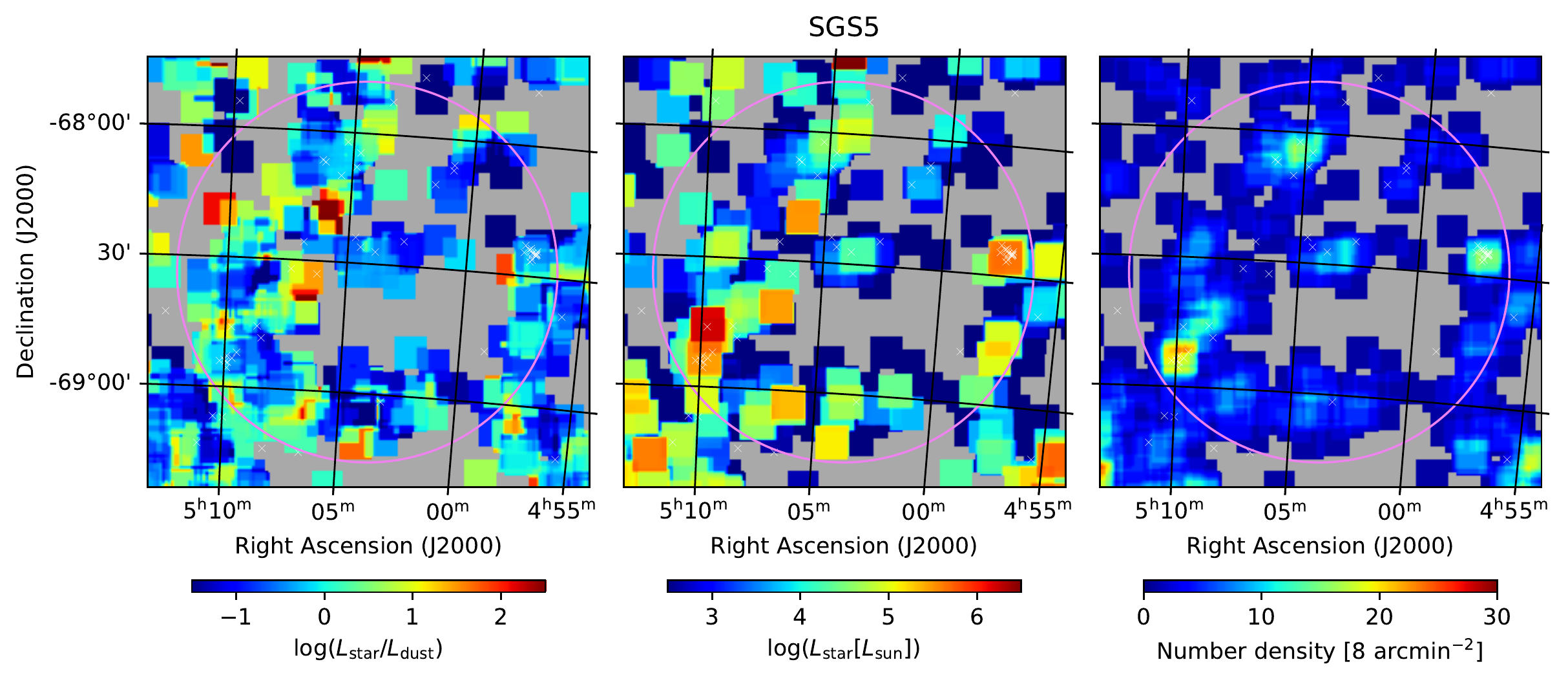}
\includegraphics[width=\textwidth]{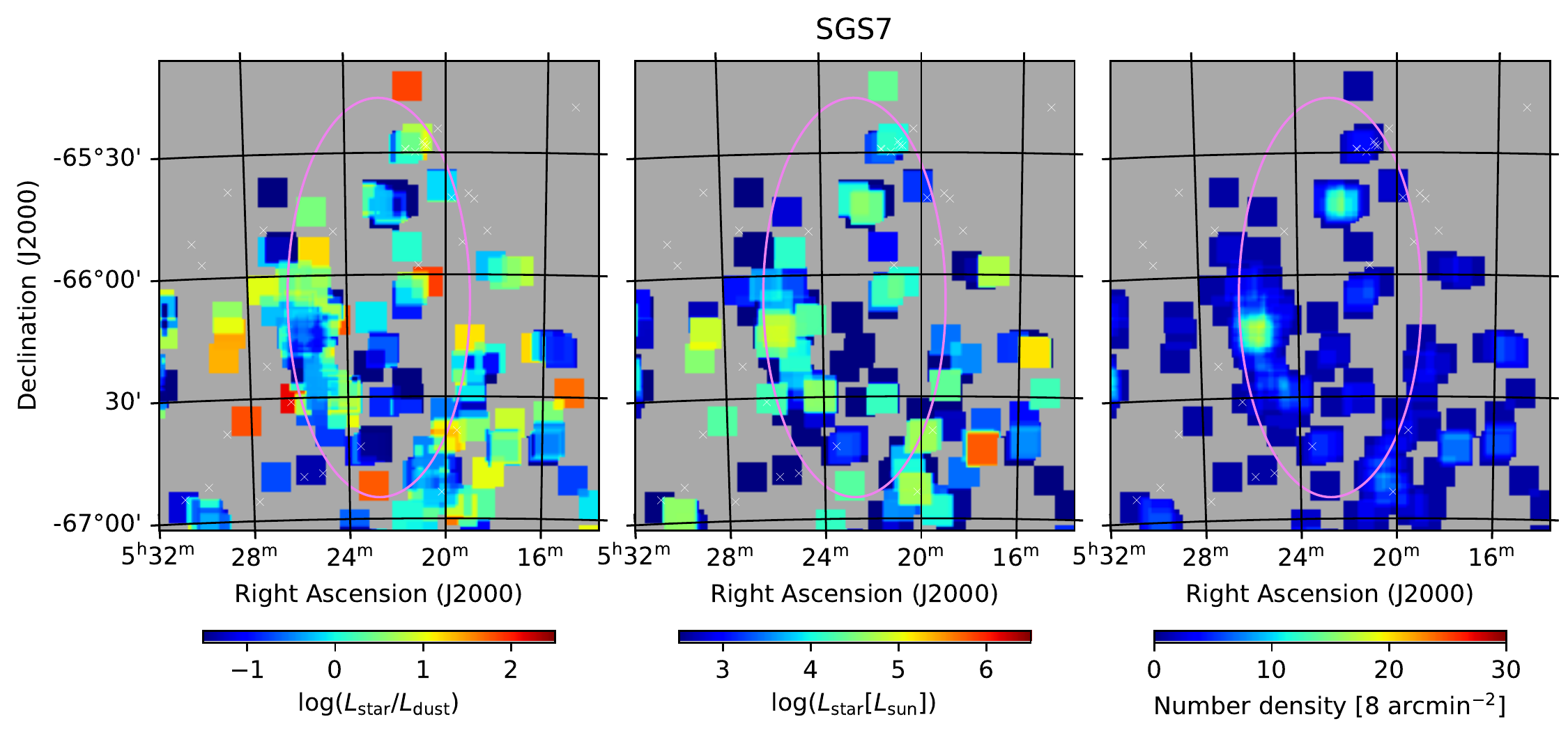}
\caption{Same as Fig.~\ref{fig:map_bin}, but close-up views of SGS5 (top) and SGS7 (bottom). Violet lines and white crosses indicate the boundaries of SGS and the positions of OB stars \citep{bon09}, respectively. \label{fig:map_bin_sgs}}
\end{figure}

Considering that the shock waves in SGSs expand outward, older YSOs are expected to be present closer to the centers of SGSs. The expansion velocities of SGSs are up to $50$~km~s$^{-1}$ \citep{boo09}, which means that the shock waves of SGSs can sweep up the ISM within a radial distance of $500$~pc in $10$~Myr. Hence older YSOs with an age of several Myr are likely to be present near the centers of SGSs, while younger YSOs in the outer shells. We investigate this age gradient by correlation analyses between  $L_\mathrm{star}/L_\mathrm{dust}$ and the radial distances from the centers of SGSs as summarized in Table~\ref{tab:sgs}. We find that only SGS12 and SGS22 show significant anti-correlations between $L_\mathrm{star}/L_\mathrm{dust}$ and the radial distance with $p=0.03$ and $p=0.04$, respectively. We speculate that recent star formation producing younger YSOs, i.e. those with lower $L_\mathrm{star}/L_\mathrm{dust}$, may have been triggered by massive stars in the inner part of SGS, which may obscure the age gradient of YSOs in the other SGSs. Figure~\ref{fig:map_bin_sgs} is the same as Fig.~\ref{fig:map_bin}, but close-up views of SGS5 and SGS7, where the positions of OB stars cataloged by \citet{bon09} are indicated. The same figures for the other SGSs are in Appendix~\ref{sec:app}. We find that $L_\mathrm{star}/L_\mathrm{dust}$ of the YSOs tends to be lower, i.e. the evolutionary stage is younger, near the massive stars in some SGSs, such as SGS5 and SGS11, supporting our scenario. On the other hand, this is not always the case for all the SGSs (e.g. SGS7). As \citet{bon09} compiled the massive star catalogs of specific regions of the LMC from the literature, their catalog is not likely to be complete. Thus we may need a more complete catalogs of massive stars to confirm our scenario.

Collisions between SGSs can promote the formation of molecular gas and stars as observed in the LMC and our Galaxy \citep{fuj14,daw15}. In particular, \citet{fuj14,fuj21} claim that collisions between SGS7 and SGS11 have triggered the formation of molecular gas and stars in the star-forming regions N48 and N49. We here investigate the function of $M_\mathrm{star}$ for our sample YSOs surrounded by more than one SGS in Fig.~\ref{fig:map_Lsd}, i.e. those likely associated with collisions between SGSs. We find that there is no significant difference in the distribution of $M_\mathrm{star}$ between the YSOs likely associated with collisions between SGSs and those with a single SGS with $p_\mathrm{KS}=0.13$. Hence, although gas compression at the interface between colliding SGSs is likely to promote star formation there \citep{fuj14,fuj21,daw15}, it may be ineffective for massive star formation, possibly due to destruction of clouds by the shocks of SGSs as discussed in Sect.~\ref{sec:dis:sf}.

\subsection{Properties of local star formation in the \ion{H}{1} ridge region} \label{sec:dis:smc}
% HI ridge region
\begin{figure}[t]
\includegraphics[width=\textwidth]{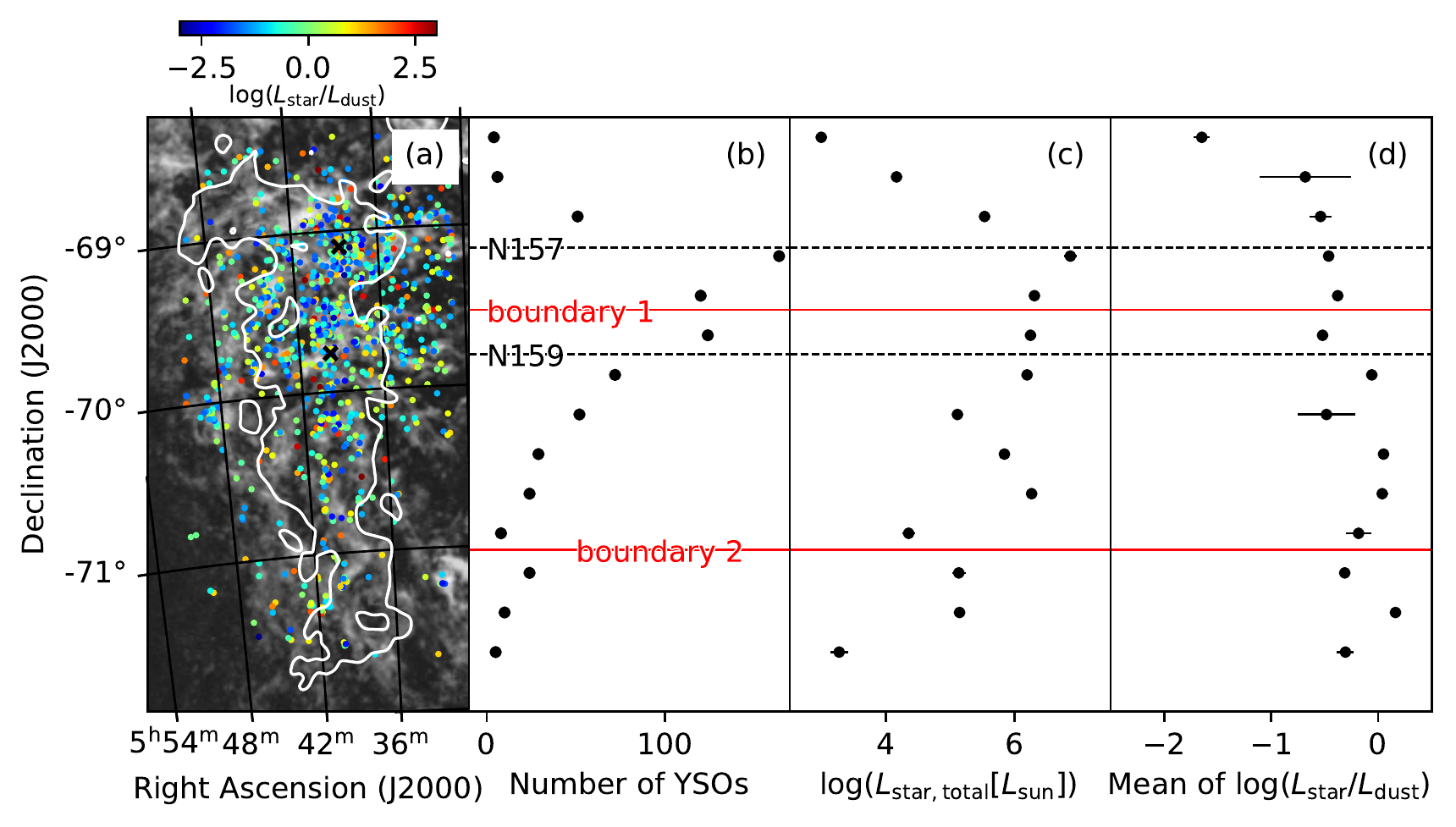}
\caption{(a) Spatial distributions of well-fitted YSOs overlaid on the \ion{H}{1} peak brightness temperature map \citep{kim99} for the \ion{H}{1} ridge region. The gray scale is in a range from $0$ to $100$~K in the brightness temperature. Data points are color-coded as in Fig.~\ref{fig:map_Lsd}. Black crosses show the positions of N157 and N159. Contours represent the distribution of hydrogen gas for the intermediate velocity component with the total hydrogen column density of $1~{\times}~10^{21}$~cm$^{-2}$ \citep{tsu20}. (b) Number of YSOs inside the contours shown in panel (a), estimated over right ascension in steps of $15{\arcsec}$ for declination. Dotted black lines show the positions of N157 and N159, while solid red lines show boundaries 1 (top) and 2 (bottom) defined in \citet{fur21}. (c) Same as panel (b), but for the total $L_\mathrm{star}$. (d) Same as panel (b), but for the mean of $L_\mathrm{star}/L_\mathrm{dust}$. \label{fig:map_ridge}}
\end{figure}

As described in Sect.~\ref{sec:dis:sf}, the \ion{H}{1} ridge region is thought to have experienced the gas flow induced by the tidal interaction between the LMC and the SMC \citep{fuk17,tsu20}. We here study the nature of the star formation in the \ion{H}{1} ridge region, using the properties of our sample YSOs. Figure~\ref{fig:map_ridge} shows the \ion{H}{1} ridge region together with the positions of the massive star-forming regions N157 and N159. The figure also represents boundaries 1 and 2 which are defined by \citet{fur21} to describe the three-dimensional geometry of the ISM in this region; they analyzed histograms of the dust extinction derived from the near-IR color excess of stars to infer the distributions of the ISM along the line of sight. As a result, they find that clouds flowed from outside the LMC are present in front of the galaxy disk in the northern region of boundary 1, while the clouds are mixed with the LMC disk in an intermediate region between boundaries 1 and 2. In the southern region of boundary 2, the clouds are present behind the LMC disk. Together with the line-of-sight velocity of the inflow gas, these results suggest that the clouds may flow in from the backside of the LMC, colliding with the LMC disk in the northern region prior to the southern region. This picture is likely to be consistent with the presence of N157 and N159 which show evidence for massive star formation triggered by the gas flow near boundary 1 \citep{fuk17,fuk19,tok19}.

Figure~\ref{fig:map_ridge} shows spatial distributions of the the number of the YSOs, the total $L_\mathrm{star}$ and the mean of $L_\mathrm{star}/L_\mathrm{dust}$ in the \ion{H}{1} ridge region. The figure shows that the number of YSOs and the total $L_\mathrm{star}$ increase toward N157 and N159, which is consistent with the picture that the active star formation is triggered by the gas flow. Figure~\ref{fig:map_ridge} also shows that $L_\mathrm{star}/L_\mathrm{dust}$ tends to be lower toward the northern region of boundary 1, suggesting that recent star formation around boundary 1 is likely to be in earlier evolutionary stages. This result supports the aforementioned three-dimensional geometry of the gas flow; collisions between the clouds triggering recent star formation may proceed in the northern region prior to the southern region \citep{tsu20,fur21}.

\section{Conclusion} \label{sec:con}
In order to study properties of star formation of the LMC, we investigated evolutionary stages and stellar masses of YSOs in the galaxy. We constructed our YSO sample combining the catalogs of YSOs established with the Spitzer and Herschel surveys of the galaxy. In total, our YSO sample contains $4825$ objects. The SEDs of the YSOs were fitted with a model consisting of stellar, PAH and two- or three-component dust emissions, from which we estimated $L_\mathrm{star}$ and $L_\mathrm{dust}$ for each YSO. We adopted $L_\mathrm{star}/L_\mathrm{dust}$ to study the evolutionary stages of the YSOs.

The SED fits were accepted with a confidence level of $95\%$ for $4098$ YSOs. The dust temperatures thus derived are consistent with those expected from the disks of YSOs. Older YSOs tend to be lacking in PAHs, indicating that PAHs may be destroyed and/or blown out from the central stars due to the radiation field at later evolutionary stages. We find significant correlations between $L_\mathrm{star}/L_\mathrm{dust}$ and the \ion{H}{1} peak brightness temperature, indicating that younger YSOs are associated with a larger amount of the ISM. We also find significant correlations between the total $L_\mathrm{star}$ and the dust surface brightness, indicating that many and/or massive YSOs are present in active star-forming regions.

$L_\mathrm{star}/L_\mathrm{dust}$ of the YSOs tends to be higher in N157 than in other active star-forming regions of the LMC, suggesting that recent star formation, i.e. mass evolution, in N157 is possibly in later evolutionary stages, which is consistent with the fact that N157 harbors massive clusters. We study the function of $M_\mathrm{star}$ of the YSOs, dividing them into those associated with the SGS, \ion{H}{1} ridge and field regions which are likely to possess different mechanisms to compress the ISM. We find that the SGS region has relatively bottom-heavy $M_\mathrm{star}$ function compared to the other regions, indicating that low-mass stars may be preferentially formed around SGSs and that it may be difficult for SGSs to compress the ISM enough to form massive stars. This may also be the case for the interface between colliding SGSs, because there is no significant difference in the function of $M_\mathrm{star}$ between the YSOs likely associated with collisions between SGSs and those with a single SGS.

In the \ion{H}{1} ridge region where the tidally-driven gas flow is thought to interact with the LMC disk, the number of YSOs and the total $L_\mathrm{star}$ increase toward N157 and N159. This is consistent with their active star formation likely induced by the gas flow. $L_\mathrm{star}/L_\mathrm{dust}$ tends to be lower toward the northern region, suggesting that recent star formation in the northern region is likely to be in earlier evolutionary stages. This result supports the picture that collisions between the gas flow and the LMC disk triggering recent star formation may proceed in the northern region prior to the southern region.

%% Putting eqnarrays or equations inside the mathletters environment groups
%% the enclosed equations by letter. For instance, the eqnarray below, instead
%% of being numbered, say, (4) and (5), would be numbered (4a) and (4b).
%% LaTeX the paper and look at the output to see the results.

\begin{acknowledgments}
This work is based in part on observations made with the Spitzer Space Telescope, which was operated by the Jet Propulsion Laboratory, California Institute of Technology under a contract with NASA. Herschel is an ESA space observatory with science instruments provided by European-led Principal Investigator consortia and with important participation from NASA. The IRSF project is a collaboration between Nagoya University and the SAAO supported by Grants-in-Aid for Scientific Research on Priority Areas (A) (nos. 10147207 and 10147214) and the Optical \& Near-Infrared Astronomy Inter-University Cooperation Program, from the Ministry of Education, Culture, Sports, Science and Technology (MEXT) of Japan and the National Research Foundation (NRF) of South Africa. This work was supported by JSPS KAKENHI Grant Number JP19K14757.

\end{acknowledgments}


\vspace{5mm}
\facilities{Spitzer, Herschel, IRSF}

%\facilities{HST(STIS), Swift(XRT and UVOT), AAVSO, CTIO:1.3m,


%\software{astropy \citep{2013A&A...558A..33A,2018AJ....156..123A},  
%          Cloudy \citep{2013RMxAA..49..137F}, 
%          Source Extractor \citep{1996A&AS..117..393B}
%          }

\bibliography{ref}{}
\bibliographystyle{aasjournal}



\appendix
\renewcommand{\thefigure}{\Alph{section}\arabic{figure}}
\setcounter{figure}{0}
\section{Close-up views of the $L_{\rm star}/L_{\rm dust}$, $L_{\rm star}$, and number density maps of well-fitted YSOs in Supergiant shells} \label{sec:app}

\begin{figure}[h]
\gridline{\fig{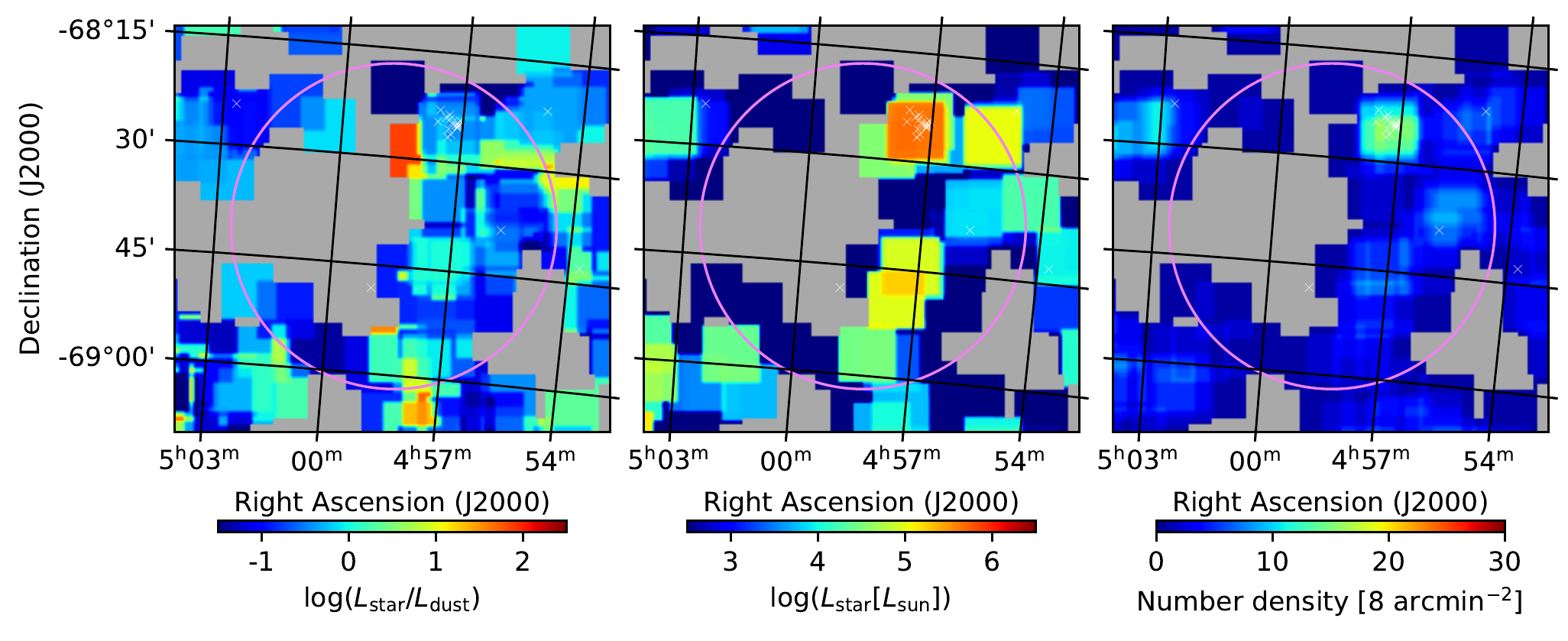}{0.5\textwidth}{SGS2}
          \fig{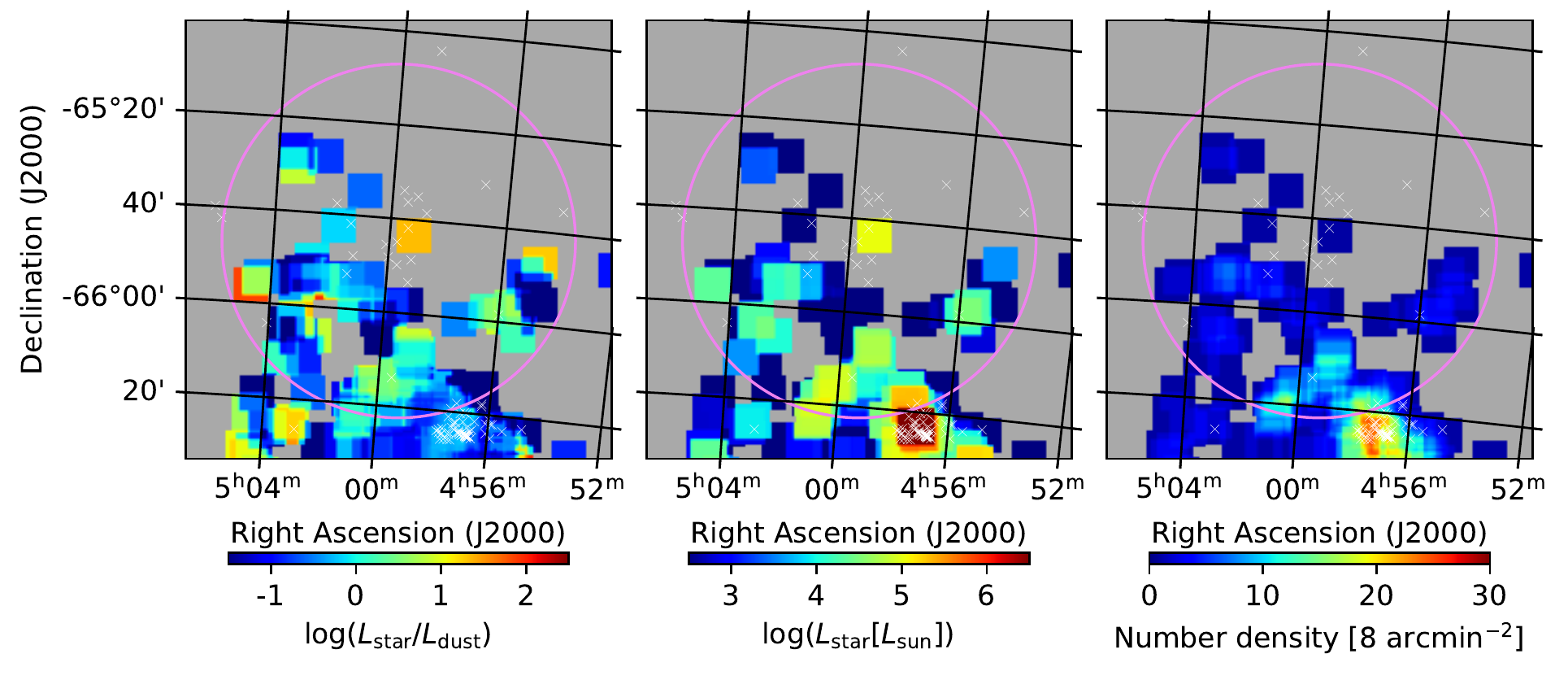}{0.5\textwidth}{SGS3}
          }
\gridline{\fig{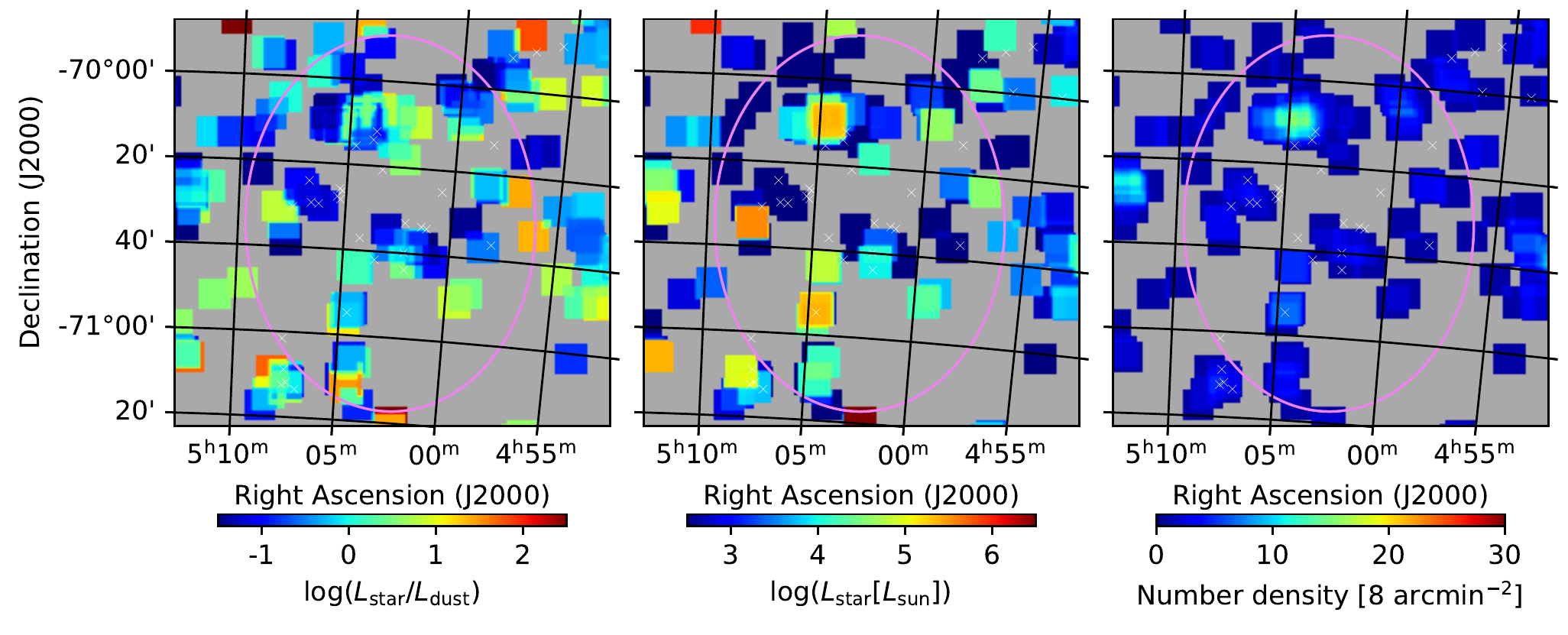}{0.5\textwidth}{SGS4}
          \fig{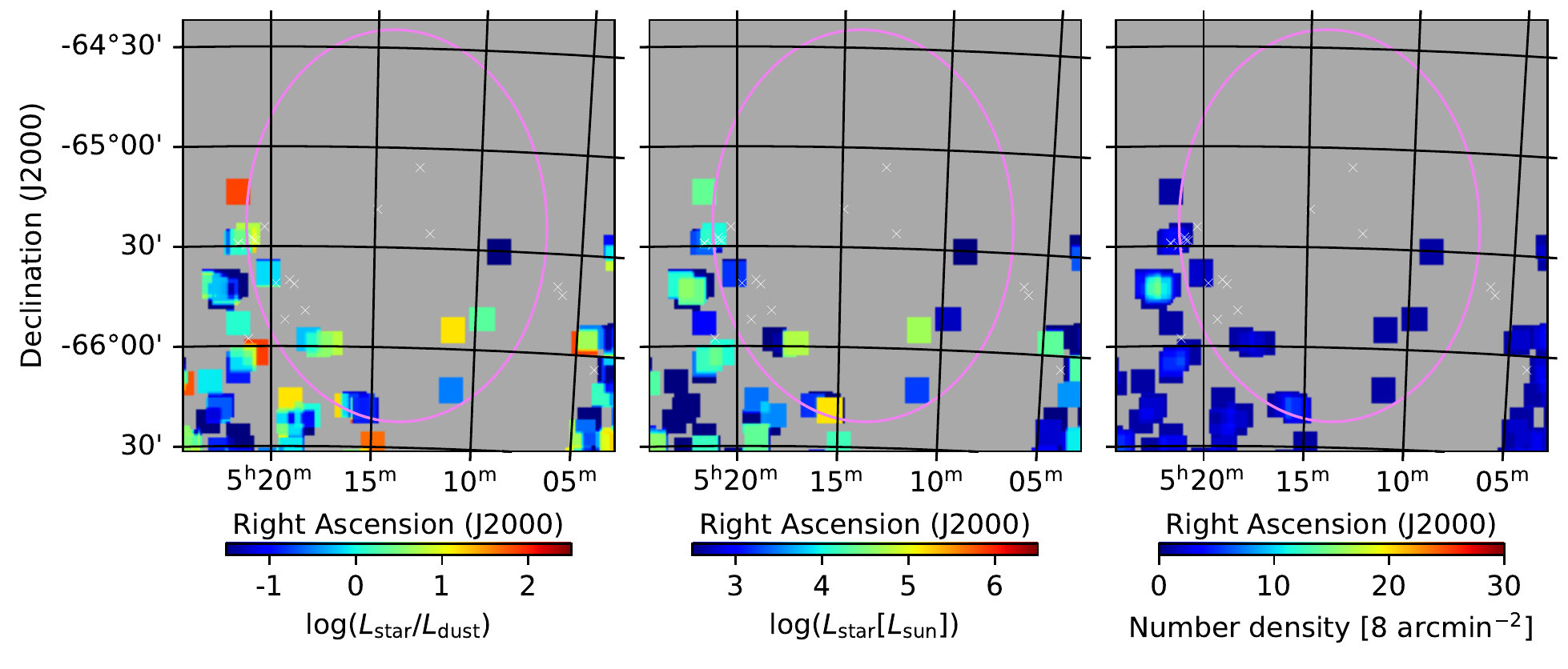}{0.5\textwidth}{SGS5}
          }
\gridline{\fig{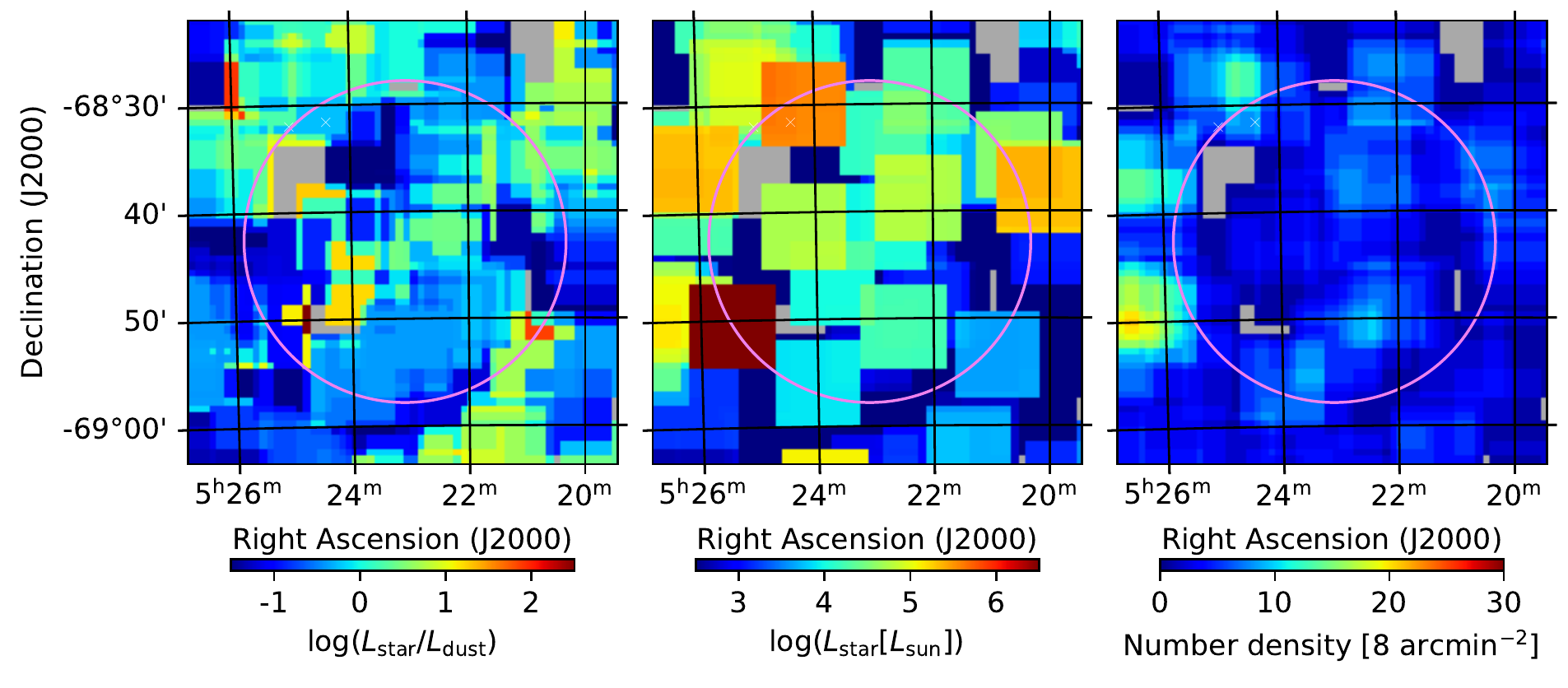}{0.5\textwidth}{SGS8}
          \fig{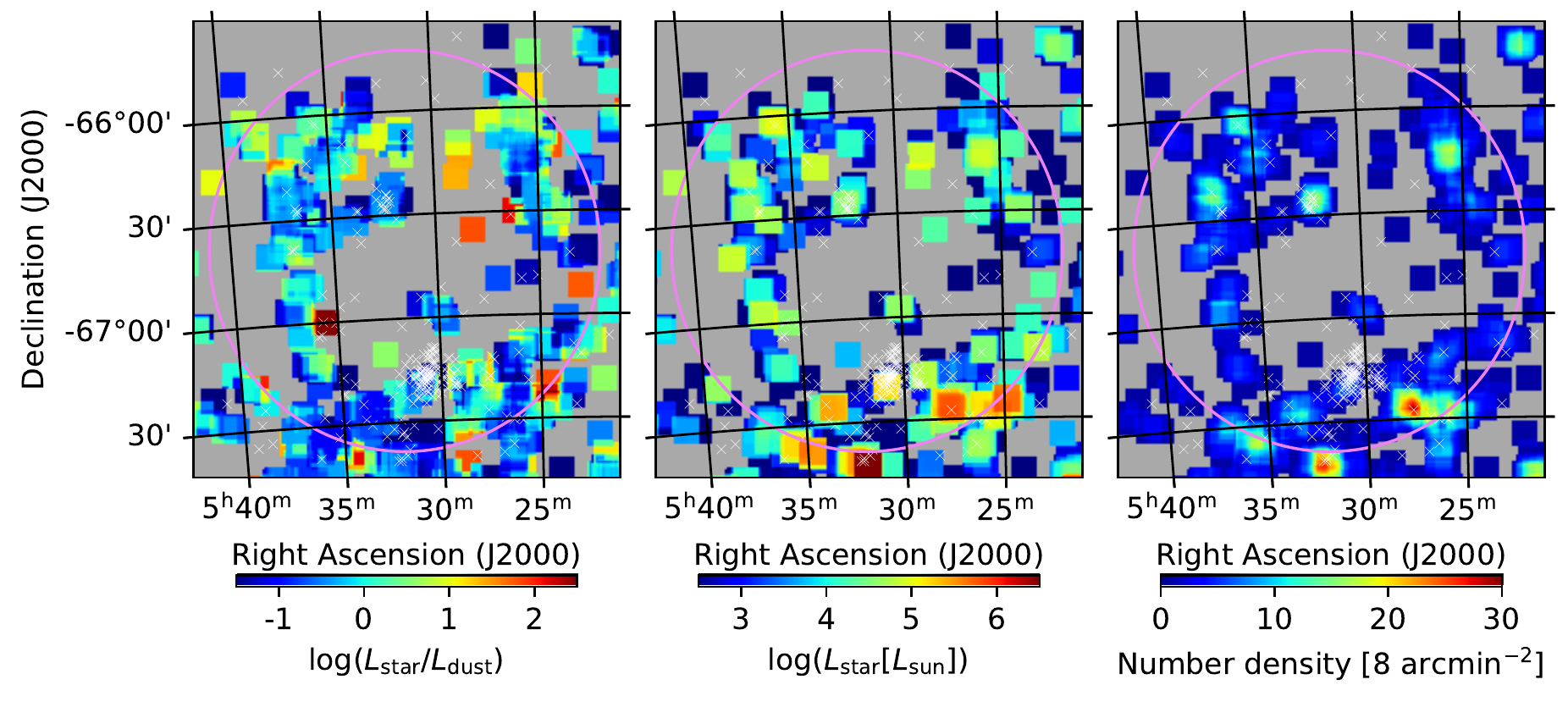}{0.5\textwidth}{SGS11}
          }
\caption{Same as Fig.~\ref{fig:map_bin_sgs}, but for SGSs other than SGS5 and SGS7.}
\label{fig:map_bin_sgs_bin}
\end{figure}

\setcounter{figure}{0}

\begin{figure}
\gridline{\fig{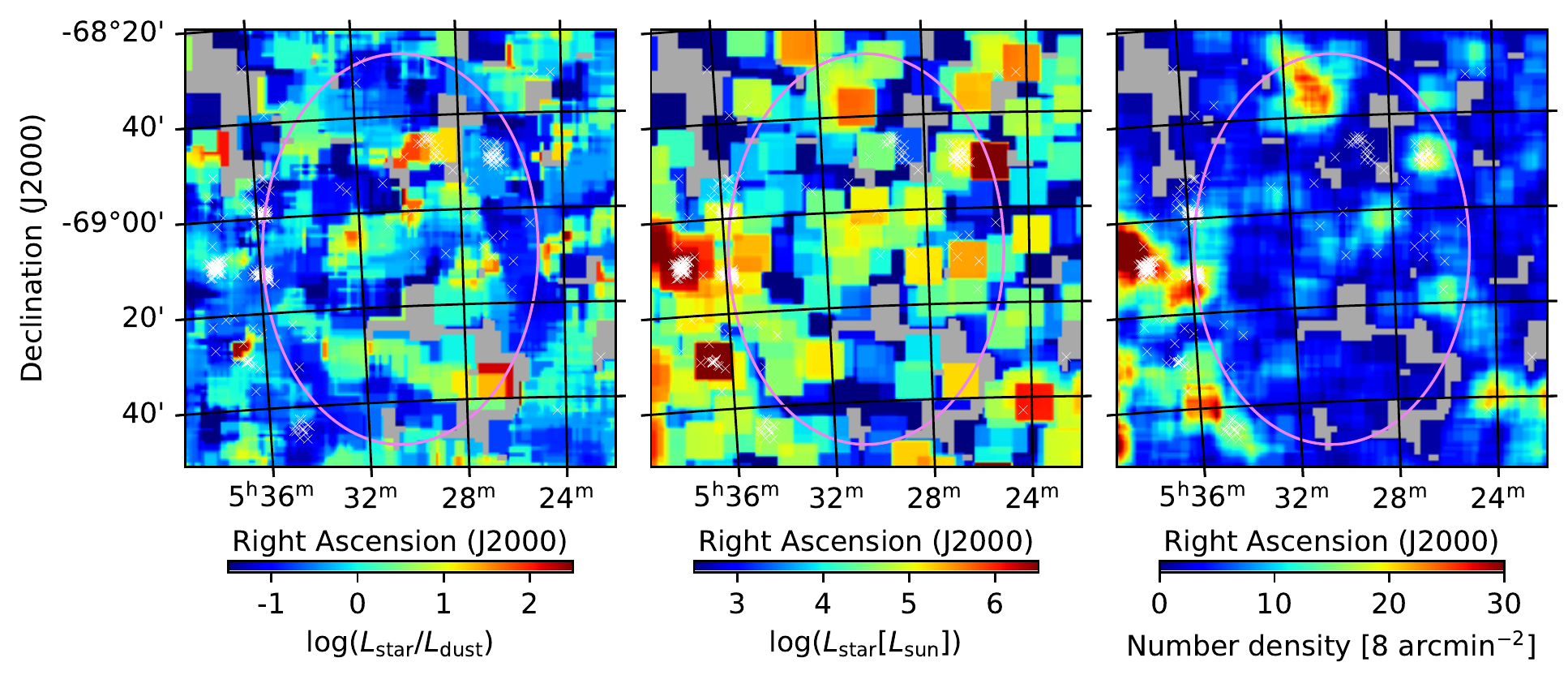}{0.5\textwidth}{SGS12}
          \fig{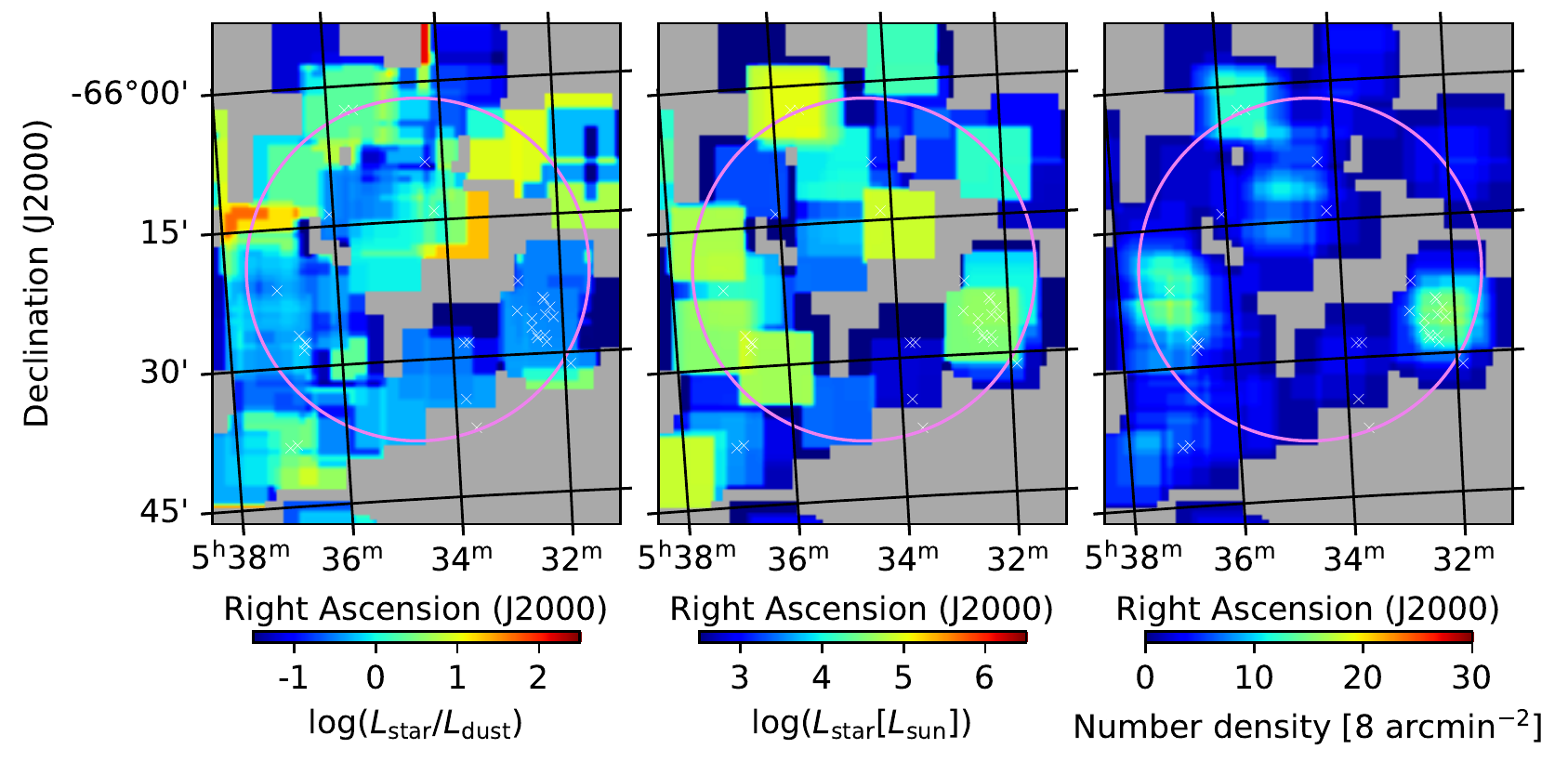}{0.5\textwidth}{SGS14}
          }
\gridline{\fig{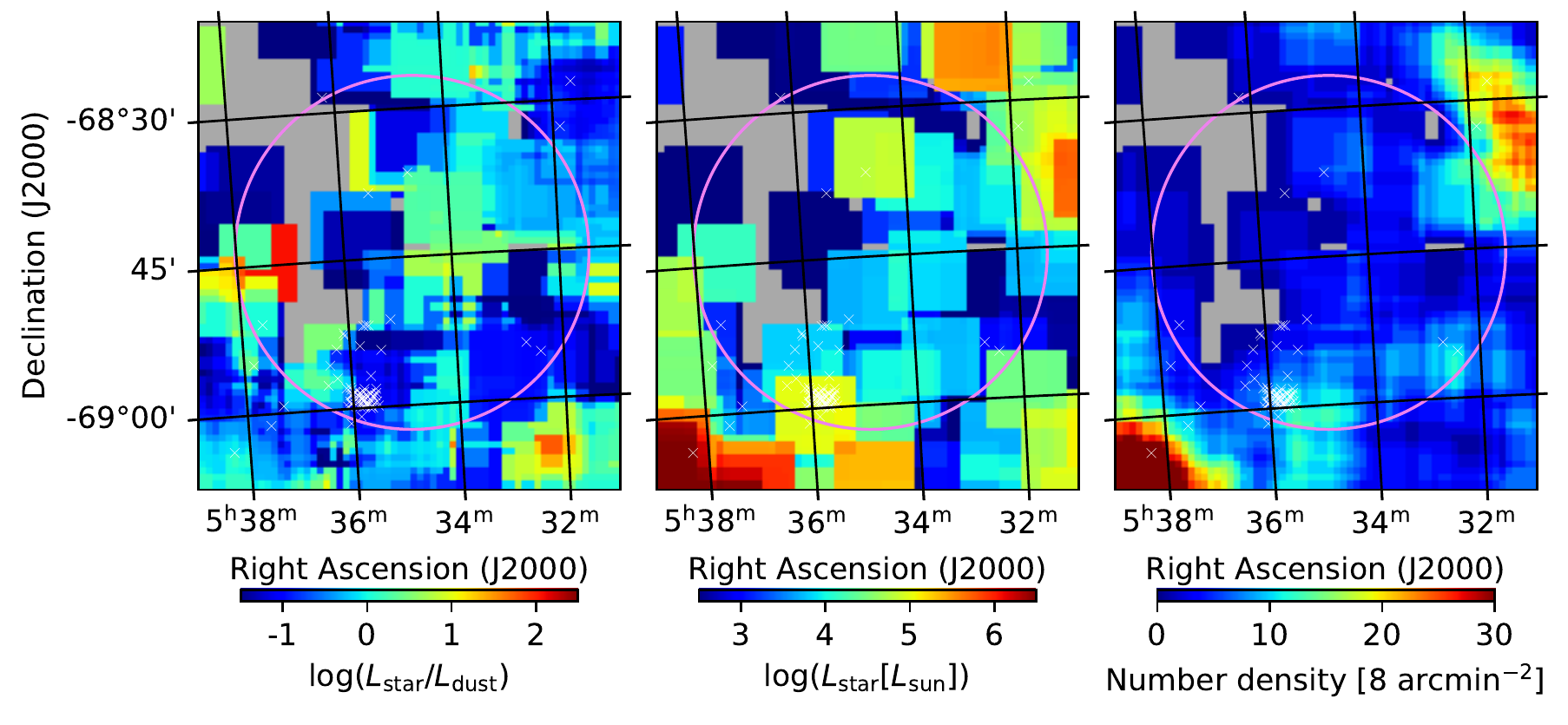}{0.5\textwidth}{SGS15}
          \fig{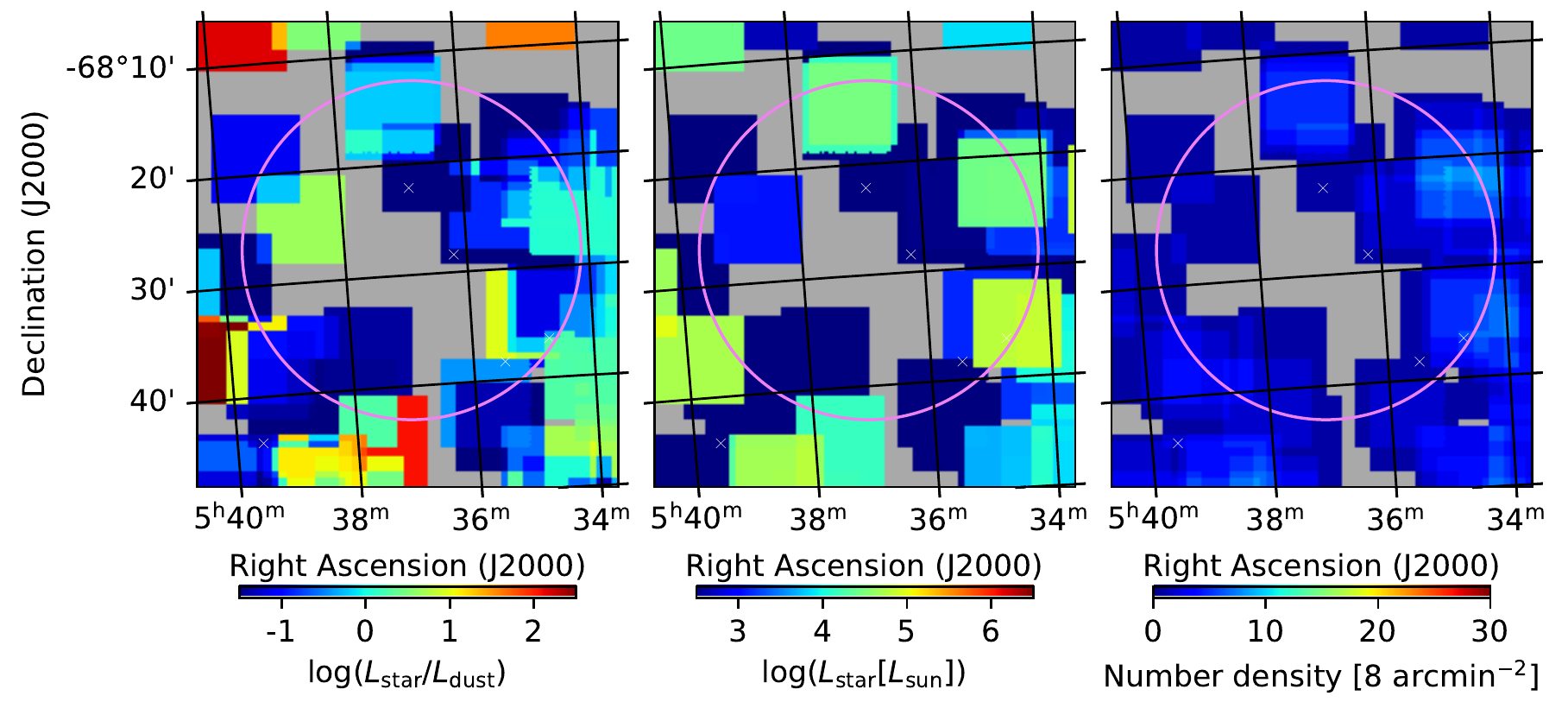}{0.5\textwidth}{SGS16}
          }
\gridline{\fig{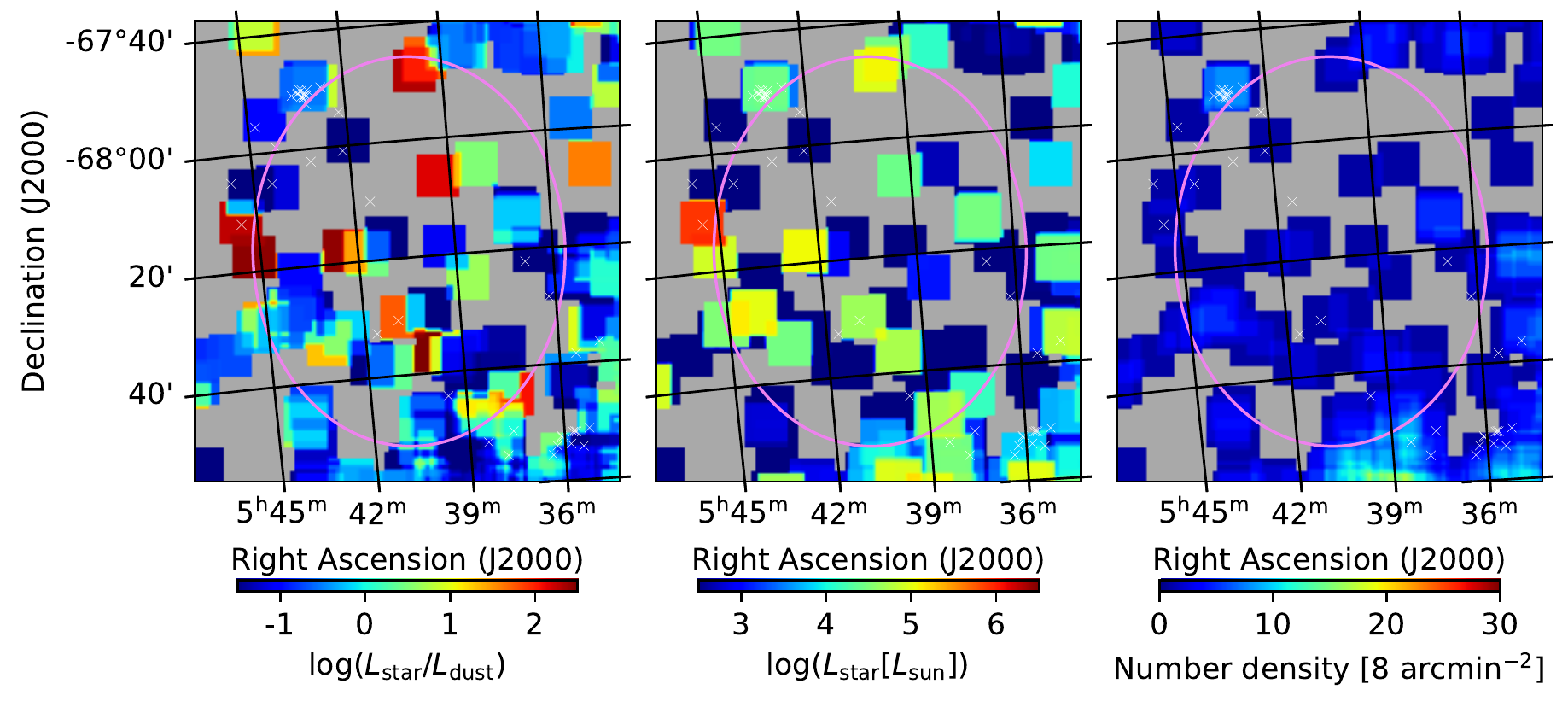}{0.5\textwidth}{SGS17}
          \fig{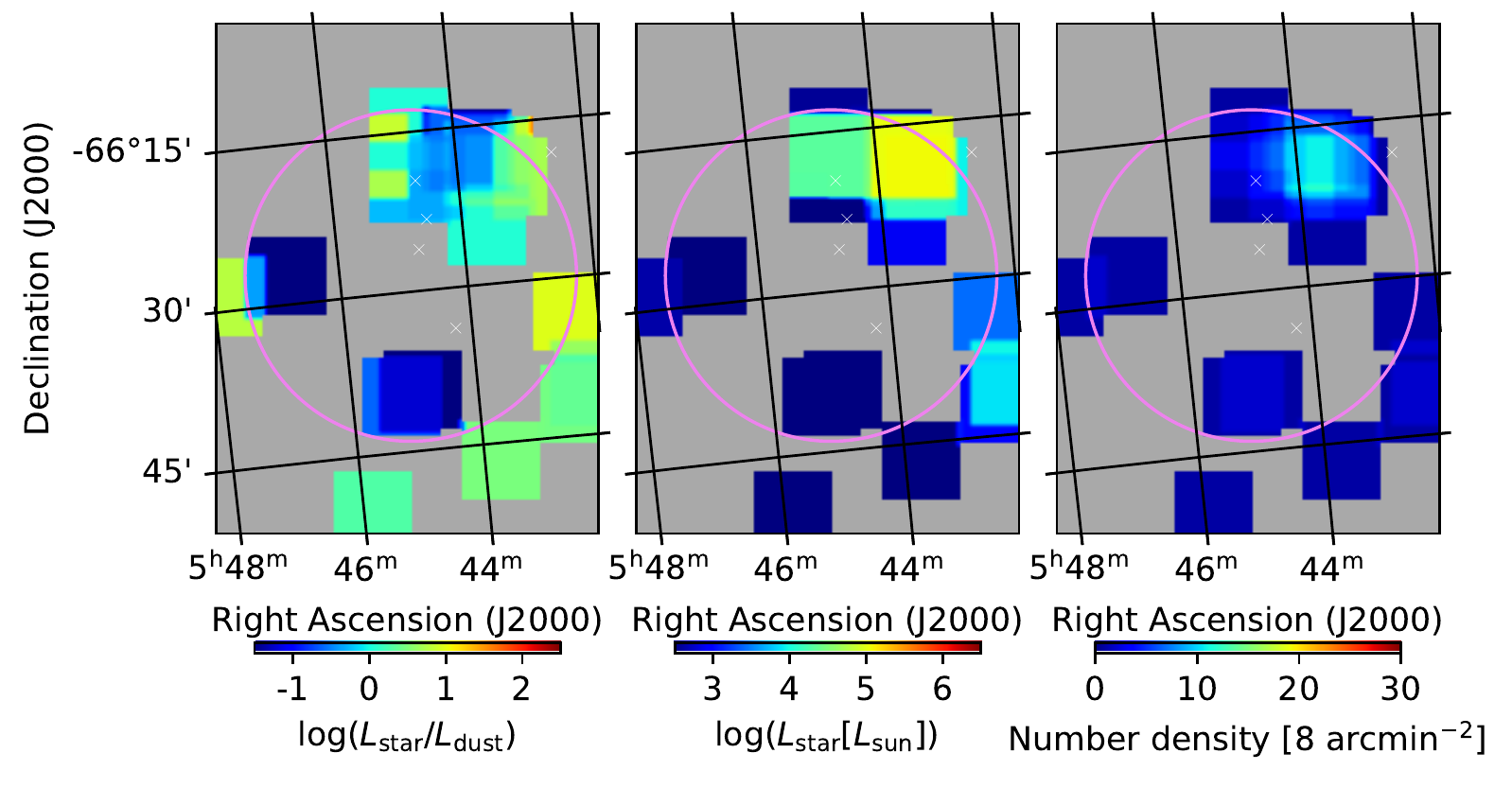}{0.5\textwidth}{SGS21}
          }
\gridline{\leftfig{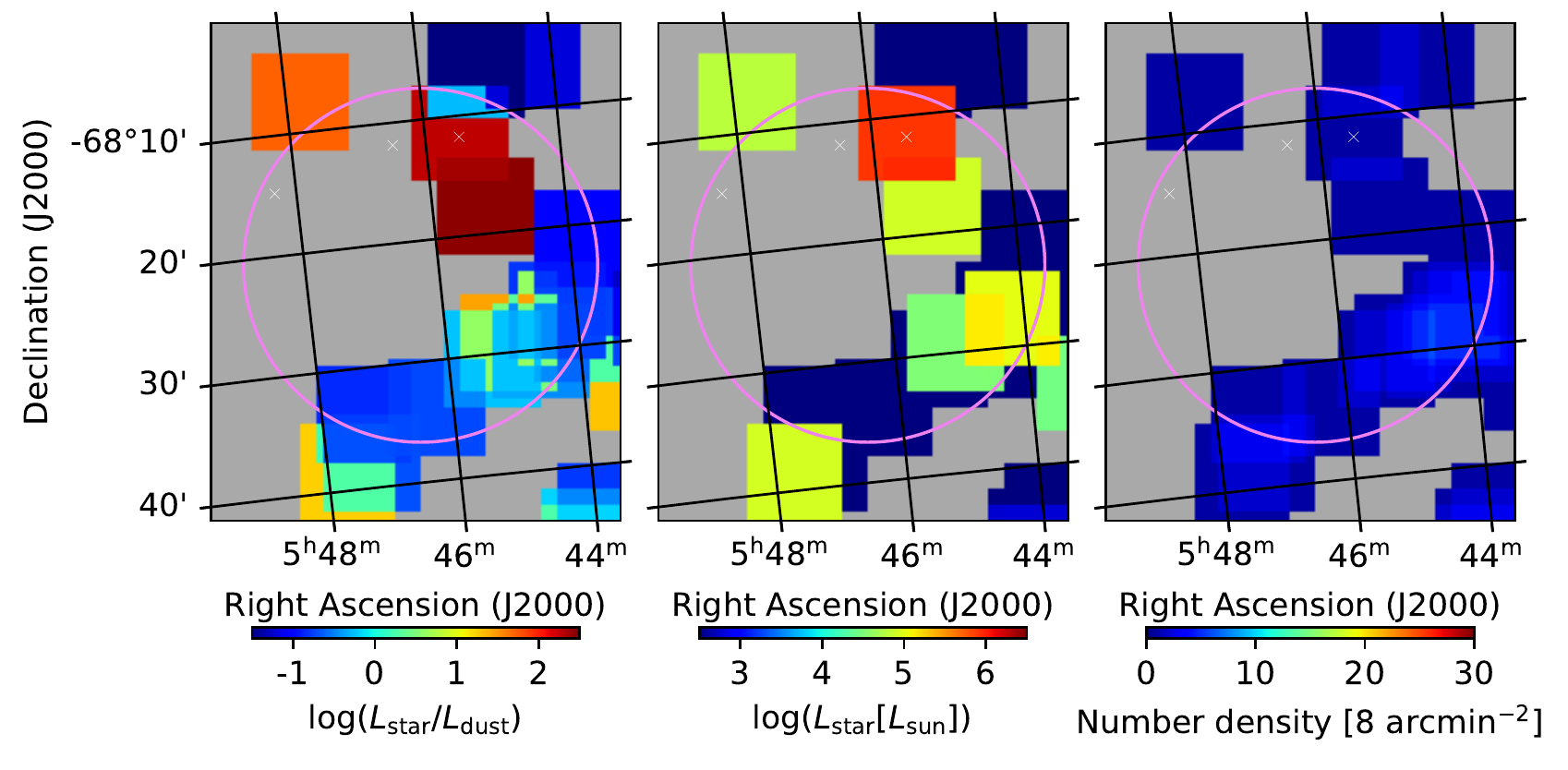}{0.5\textwidth}{SGS22}
          \fig{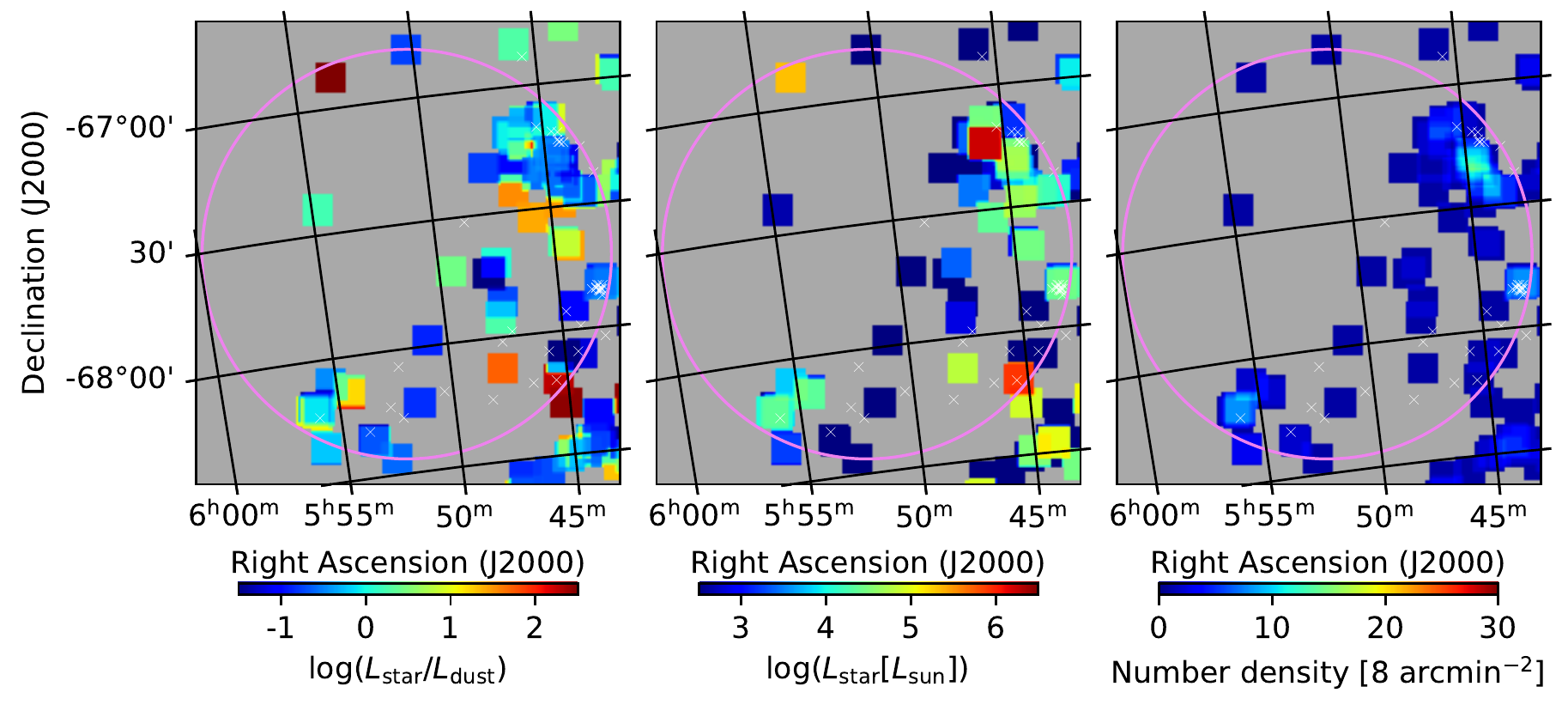}{0.5\textwidth}{SGS23}
          }
\caption{Continued.}
\end{figure}




\end{document}